\documentclass[twocolumn]{aastex6}

\DeclareGraphicsExtensions{.jpg,.pdf,.png,.eps,.ps}

\newcommand{\msun}{$M_{\sun}$}
\def\arcmin{\hbox{$^\prime$}}
\def\arcsec{\hbox{$^{\prime\prime}$}}
\def\arcdeg{\mbox{$^\circ$}}

\newcommand{\hdelta}{H-$\delta$}
\newcommand{\oii}{[O {\small II}] $\lambda$3727}

\shortauthors{Bayliss+2016}

\begin{document}

\def\MIT{1}
\def\Colby{2}
\def\Harvard{3}
\def\AAUChicago{4}
\def\KICPChicago{5}
 \def\FNAL{6}
\def\PhysicsUChicago{7}
\def\Argonne{8}
\def\Miss{9}
\def\Munich{10}
\def\ExcellenceCluster{11}
\def\Colorado{12}
\def\NASAAmes{13}
\def\Hawaii{14}
\def\CfA{15}
\def\Sinica{16}
\def\CTIO{17}

\altaffiltext{\MIT}{Kavli Institute for Astrophysics \& Space Research, Massachusetts Institute of Technology, 77 Massachusetts Ave., 
Cambridge, MA 02139, USA}
\altaffiltext{\Colby}{Department of Physics and Astronomy, Colby College, 5100 Mayflower Hill Dr, Waterville, ME 04901, USA}
\altaffiltext{\Harvard}{Department of Physics, Harvard University, 17 Oxford Street, Cambridge, MA 02138, USA}
\altaffiltext{\AAUChicago}{Department of Astronomy and Astrophysics, University of Chicago, Chicago, IL, USA 60637, USA}
\altaffiltext{\KICPChicago}{Kavli Institute for Cosmological Physics, University of Chicago, Chicago, IL, USA 60637, USA}
\altaffiltext{\FNAL}{Fermi National Accelerator Laboratory, Batavia, IL 60510-0500, USA}
\altaffiltext{\PhysicsUChicago}{Department of Physics, University of Chicago, Chicago, IL, USA 60637, USA}
\altaffiltext{\Argonne}{Argonne National Laboratory, Argonne, IL, USA 60439, USA}
\altaffiltext{\Miss}{Department of Physics and Astronomy, University of Missouri, 5110 Rockhill Road, Kansas City, MO 64110, USA}
\altaffiltext{\Munich}{Faculty of Physics, Ludwig-Maximilians-Universit\"{a}t, Scheinerstr.\ 1, 81679 Munich, Germany}
\altaffiltext{\ExcellenceCluster}{Excellence Cluster Universe, Boltzmannstr.\ 2, 85748 Garching, Germany}
\altaffiltext{\Colorado}{Center for Astrophysics and Space Astronomy, Department of Astrophysical and Planetary Science, University of Colorado, Boulder, C0 80309, USA}
\altaffiltext{\NASAAmes}{NASA Ames Research Center, Moffett Field, CA 94035, USA}
\altaffiltext{\Hawaii}{Institute for Astronomy, the University of Hawaii, 2680 Woodlawn Dr, Honolulu, HI 96822, USA}
\altaffiltext{\CfA}{Harvard-Smithsonian Center for Astrophysics, 60 Garden Street, Cambridge, MA 02138, USA}
\altaffiltext{\Sinica}{Academia Sinica Institute of Astronomy and Astrophysics (ASIAA) 11F of AS/NTU Astronomy-Mathematics Building,No.1, Sec. 4, Roosevelt Rd, Taipei 10617, Taiwan}
\altaffiltext{\CTIO}{Cerro Tololo Inter-American Observatory, Casilla 603, La Serena, Chile}

\title{Velocity Segregation and Systematic Biases In Velocity Dispersion Estimates With the SPT-GMOS Spectroscopic Survey}

\author{Matthew.~B.~Bayliss\altaffilmark{\MIT}, 
Kyle~Zengo\altaffilmark{\Colby}, 
Jonathan~Ruel\altaffilmark{\Harvard}, 
Bradford~A.~Benson\altaffilmark{\AAUChicago,\KICPChicago,\FNAL}, 
Lindsey~E.~Bleem\altaffilmark{\KICPChicago,\PhysicsUChicago,\Argonne}, 
Sebastian~Bocquet\altaffilmark{\KICPChicago,\Argonne}, 
Esra~Bulbul\altaffilmark{\MIT}, 
Mark~Brodwin\altaffilmark{\Miss}, 
Raffaella~Capasso\altaffilmark{\Munich,\ExcellenceCluster}, 
I-non~Chiu\altaffilmark{\Sinica}, 
Michael~McDonald\altaffilmark{\MIT}, 
David Rapetti\altaffilmark{\Colorado,\NASAAmes},
Alex Saro\altaffilmark{\Munich,\ExcellenceCluster}, 
Brian~Stalder\altaffilmark{\Hawaii}, 
Antony~A.~Stark\altaffilmark{\CfA}, 
Veronica~Strazzullo\altaffilmark{\Munich,\ExcellenceCluster}, 
Christopher~W.~Stubbs\altaffilmark{\Harvard,\CfA}, 
Alfredo~Zenteno\altaffilmark{\CTIO}}

\email{mbayliss@mit.edu}

\begin{abstract}

The velocity distribution of galaxies in clusters is not universal; rather, galaxies are segregated according 
to their spectral type and relative luminosity. We examine the velocity distributions of different populations 
of galaxies within 89 Sunyaev Zel'dovich (SZ) selected galaxy clusters spanning $ 0.28 < z < 1.08$. Our 
sample is primarily draw from the SPT-GMOS spectroscopic survey, supplemented by additional published 
spectroscopy, resulting in a final spectroscopic sample of 4148 galaxy spectra---2868 cluster members. 
The velocity dispersion of star-forming cluster galaxies is $17\pm4$\% greater than that of passive cluster 
galaxies, and the velocity dispersion of bright ($m < m^{*}-0.5$) cluster galaxies is $11\pm4$\% lower than 
the velocity dispersion of our total member population. We find good agreement with simulations regarding the 
shape of the relationship between the measured velocity dispersion and the fraction of passive vs. star-forming 
galaxies used to measure it, but we find a small offset between this relationship as measured in data 
and simulations in which suggests that our dispersions are systematically low by as much as 3\% relative 
to simulations. We argue that this offset could be interpreted as a measurement of the effective velocity bias that 
describes the ratio of our observed velocity dispersions and the intrinsic velocity dispersion of dark matter 
particles in a published simulation result. Measuring velocity bias in this way suggests that large spectroscopic 
surveys can improve dispersion-based mass-observable scaling relations for cosmology even in the face of 
velocity biases, by quantifying and ultimately calibrating them out.

\end{abstract}


\section{Introduction}

Spectroscopic surveys of massive galaxy clusters---the most massive gravitationally bound 
structures in the universe---provide important constraints on cosmological measurements 
of the growth of structure and valuable insights into the properties of galaxy populations in 
the most extreme over-dense environments. There is a long and rich history of studies using 
the widths of the peculiar velocity distribution of clusters, i.e., their velocity dispersions, to estimate 
the depths of their gravitational potential wells and infer the total masses of those clusters 
\citep{zwicky37,bahcall81,kent82,bahcall91,biviano93,girardi93,girardi96,dressler99,rines03,geller13,rines13,sifon13,ruel14,sifon16}. 
Velocity dispersion measurements are one of several mass-observable proxies that can contribute to 
calibrating mass measurements of clusters, which remains a prime concern for extracting competitive 
cosmological constraints from galaxy cluster abundance measurements 
\citep{majumdar03,majumdar04,rozo10,allen11,williamson11,benson13,planck13-XX,vonderLinden14,bocquet15,dehaan16}. 

It is also well-established that the phase space properties of galaxies within clusters are not uniform 
across all galaxy types. For example, numerous studies have observed differences in the spatial 
distribution of different types of galaxies, with red/early-type/passive galaxies preferentially occupying 
more central radial regions of clusters while blue/late-type/star-forming galaxies are more likely to 
reside at larger radii \citep{melnick77,dressler80,whitmore93,abraham96,balogh97,dressler99,balogh00,dominguez01,gerken04,rosati14}, 
at least at $z<1$ \citep{brodwin13}, and with the caveat that SZ-based cluster centroids can be 
significant, especially at $z>1$ \citep{song12}.

A similar effect is expected to occur in line-of-sight velocity space where the distribution of peculiar velocities 
of cluster galaxies are segregated by galaxy type. These velocity segregation effects are more challenging 
to measure than their spatial counterparts, but differences in the peculiar velocity distributions of different types 
of galaxies have been previously noted in studies. Most studies have found that blue/late-type/star-forming 
galaxies tend to have larger peculiar velocities than red/early-type/passive galaxies \citep{sodre89,zabludoff93,colless96,biviano97,carlberg97b,dressler99,biviano02,ribeiro10,girardi15,barsanti16,biviano16}, 
while a few have detected evidence for more luminous cluster galaxies having systematically lower peculiar 
velocities than fainter cluster galaxies \citep{chincarini77,biviano92,mohr96,goto05,ribeiro10,old2013,barsanti16}. Some 
studies, however, have detected no significant velocity segregation in galaxy cluster samples \citep{tammann72,moss77,rines03,biviano09,rines13,crawford14}, suggesting the need for more analyses 
of larger datasets. In recent work on this topic \citet{barsanti16} argue specifically for the importance of 
analyzing large spectroscopic datasets as stacks or ensembles in order to tease out velocity segregation 
effects. Analyses of large ensemble datasets are ideal for revealing the average velocity segregation effects 
associated with sub-populations of cluster member galaxies.

\begin{figure}[t]
\centering
\includegraphics[scale=0.44]{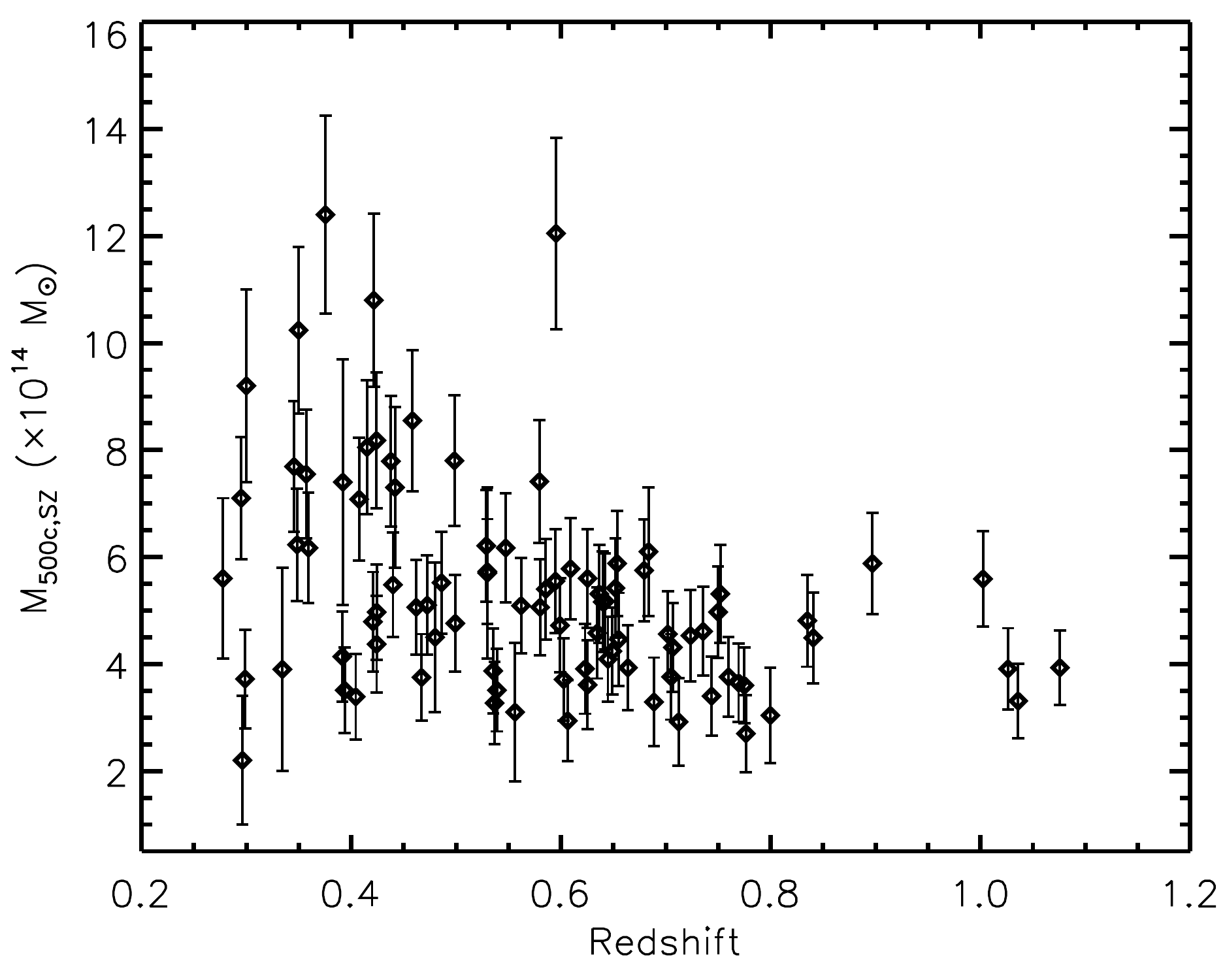}
\caption{\scriptsize{
The redshift distribution and SZ-derived masses of the galaxy clusters used in this work, 
all SZ-selected. Masses are taken from \citet{bleem15} and \citet{hasselfield2013}.}}
\label{fig:mzdist}
\end{figure}

In this work we present an analysis of a large sample of spectroscopically observed galaxies in 89 
clusters that were selected via the Sunyaev Zel'dovich effect 
\citep[SZ effect;][]{sunyaev72,sunyaev80}. Most of these clusters are from the South Pole Telescope 
(SPT) SZ survey \citep{staniszewski09,vanderlinde10,carlstrom11,williamson11,reichardt13,bleem15}, 
with additional clusters drawn from the Atacama Cosmology Telescope (ACT) SZ survey 
\citep{marriage11b,hasselfield2013}. SPT and ACT have both performed surveys optimized 
($\sim$ 1\arcmin\ beams) for identifying galaxy clusters above approximately flat mass thresholds 
at $z \gtrsim 0.25$. 

Our objective is to test for the presence of systematic biases that affect velocity dispersion estimates as 
a function of the properties of the galaxies used to estimate those dispersions. Recent simulations 
provide theoretical expectations for the scaling relation between galaxy cluster mass and 
the observed line-of-sight velocity dispersion \citep{evrard08,white10,diemer13,munari13}, but there is 
still a significant leap that must be made to robustly connect the dark matter particle dispersions that are 
easily measurable in simulations to the galaxy velocity dispersion that we actually observe 
\citep{gifford13a,saro13,hernandez14,sifon16}. In this paper we aim to quantify velocity segregation 
effects in a uniformly selected sample of massive galaxy clusters, and to compare our results to 
recent efforts to calibrate for these effects in simulations. Quantifying the degree to which velocity 
segregation effects induce systematic biases in velocity dispersions estimates would 
provide valuable input for calibrating mass-observable scaling relations in cosmological analyses of 
galaxy cluster samples that utilize velocity dispersion measurements \citep{borgani97}. 

This paper is organized into the following sections: in \S~\ref{sec:specdata} we describe the 
dataset used in our analysis, including new spectroscopy as well as spectra taken from the literature. 
In \S~\ref{sec:vse} we describe the process that we use to generate an ensemble cluster galaxy velocity 
sample, and we then use that ensemble to measure velocity segregation as a function of spectral type 
and relative luminosity. In \S~\ref{sec:biases} 
we compare our velocity segregation measurements to similar measurements in simulated galaxy 
clusters to recover estimates of the systematic biases between velocity dispersions measured in real 
data vs simulations. In \S~\ref{sec:discussion} we discuss the results of our analysis and explore 
the possible implications. Finally, in \S~\ref{sec:conclusion} we summarize our results and conclusions. 
Unless explicitly stated otherwise, the uncertainties that we report reflect 68\% confidence intervals. 
All magnitude measurements throughout this paper are in the AB system, and for all cosmological 
computations we assume a flat cosmology with $\Omega_{M} = 0.3$ 
and $\Omega_{\Lambda} = 0.7$, $H_{0} = 70$ km s$^{-1}$ Mpc$^{-1}$, and $h=H_{0}/100= 0.7$.

\def\arraystretch{0.92}
\begin{deluxetable*}{ccccccccc}
\tablecaption{Example Catalog Data for Individual Galaxies\label{tab:catalog}}
\tabletypesize{\scriptsize}
\tablehead{
\colhead{Cluster } &
\colhead{Object } &
\colhead{ RA } &
\colhead{ Dec  } &
\colhead{ z ($\delta_{z}$)\tablenotemark{a} } &
\colhead{ $W_{0}$ [O {\textsc II}] } & 
\colhead{ $W_{0}$ H-$\delta$ } &
\colhead{ $r$} &
\colhead{ $i$} \\
\colhead{ Name} &
\colhead{ Name} &
\colhead{ ($^{\circ}$) } &
\colhead{ ($^{\circ}$) } &
\colhead{  } &
\colhead{ (\AA) } &
\colhead{ (\AA) } &
\colhead{ (AB) } &
\colhead{ (AB) } } 
\startdata
SPT-CLJ2337-5942 & J233729.84-594159.0 & 354.37433 & -59.69973 & $0.77910(32)$ & $  4.50\pm 2.42$ & $  3.51\pm 1.62$ &  22.63 & 21.31  \\
SPT-CLJ2337-5942 & J233732.53-594305.9 & 354.38556 & -59.71831 & $0.77370(29)$ & $ -1.16\pm 3.93$ & $  5.62\pm 2.23$ &  22.47 & 21.39  \\
SPT-CLJ2337-5942 & J233722.38-594318.3 & 354.34326 & -59.72176 & $0.77110(44)$ & $  6.85\pm 4.30$ & $ -0.38\pm 2.77$ &  22.63 & 21.25  \\
SPT-CLJ2337-5942 & J233718.74-594215.5 & 354.32806 & -59.70430 & $0.78330(33)$ & $  5.43\pm 4.08$ & $  1.33\pm 2.72$ &  23.10 & 21.83  \\
SPT-CLJ2337-5942 & J233729.45-594257.2 & 354.37271 & -59.71588 & $0.77590(27)$ & $ -0.59\pm 4.41$ & $  1.02\pm 2.58$ &  21.86 & 20.68  \\
SPT-CLJ2337-5942 & J233725.66-594157.9 & 354.35690 & -59.69942 & $0.77550(30)$ & $ -0.02\pm 3.78$ & $  5.54\pm 2.14$ &  22.39 & 21.12  \\
SPT-CLJ2337-5942 & J233727.50-594234.7 & 354.36456 & -59.70963 & $0.77760(24)$ & $  3.60\pm 2.65$ & $  3.37\pm 1.65$ &  21.09 & 20.10  \\
SPT-CLJ2337-5942 & J233724.13-594240.6 & 354.35052 & -59.71128 & $0.78360(41)$ & $  2.26\pm 5.74$ & $  2.55\pm 3.88$ &  23.40 & 22.04  \\
SPT-CLJ2337-5942 & J233734.02-594231.9 & 354.39175 & -59.70886 & $0.77630(37)$ & $ -2.27\pm 5.45$ & $ -1.82\pm 3.09$ &  22.96 & 21.76  \\
SPT-CLJ2337-5942 & J233738.77-594312.1 & 354.41156 & -59.72004 & $0.77480(55)$ & $ -4.57\pm12.78$ & $  7.59\pm 6.92$ &  24.52 & ~23.08  
\enddata
\tablenotetext{a}{The measured redshift with the uncertainty in the last two digits given in parentheses.}
\end{deluxetable*}

\section{Spectroscopic Data and Galaxy Catalogs} 
 \label{sec:specdata}

\subsection{SPT-GMOS and Supplemental Literature Spectroscopy}

The majority of the dataset used in this analysis comes from the SPT-GMOS spectroscopic survey 
\citep{bayliss16}, which consists of spectroscopic follow-up of 62 galaxy clusters from the SPT-SZ survey. 
The full SPT-GMOS sample includes 2243 galaxy spectra, 1579 of which are cluster member galaxies. 
In addition to primary catalog information such as redshifts and positions of individual galaxies with spectra, 
the SPT-GMOS dataset also includes spectral index measurements made using the 1-dimensional (1D) 
spectrum of each individual galaxy. Where possible we supplement the SPT-GMOS spectroscopic sample 
with other published spectroscopy of SPT and ACT SZ-selected galaxy clusters 
\citep{brodwin10,sifon13,stalder13,bayliss14c,ruel14,sifon16}.

Including spectroscopy from the literature in our analysis requires access to catalog information---i.e., redshifts, 
positions---as well as the extracted 1D spectra for individual galaxies from which we measure 
specific spectral index values as the strength of specific spectral features. These spectral index measurements 
provide significant added value to the resulting galaxy spectroscopy catalog; see \S~\ref{sec:classification} 
below for a description of the spectral index measurements and the galaxy classification metrics that we use.
In total we incorporate literature spectra for an additional 27 galaxy clusters, adding 1431 galaxies 
and 1118 cluster member galaxies for which we have access to both catalogs and the extracted 
spectra. 

\begin{deluxetable*}{lccccccc}
\tablecaption{SZ-Selected Spectroscopic Cluster Sample\label{tab:spectable}}
\tabletypesize{\small}
\tablehead{
\colhead{Cluster Name} &
\colhead{N$_\mathrm{spec}$} &
\colhead{N$_\mathrm{members}$} &
\colhead{N$_\mathrm{pass+PSB}$} &
\colhead{N$_\mathrm{star-forming}$} &
\colhead{ ${\bar z_\mathrm{cluster}}$} &
\colhead{ $\sigma_\mathrm{v}$ } &
\colhead{ Reference(s)\tablenotemark{a} } \\
\colhead{  } &
\colhead{  } &
\colhead{  } &
\colhead{  } &
\colhead{  } &
\colhead{  } &
\colhead{ (km s$^{-1}$) } &
\colhead{  } }
\startdata
SPT-CLJ0000-5748 &  97 &  56 & 48 & 8 &  $0.7004\pm0.0011$ &  $682\pm 108$ &  2, 4  \\
SPT-CLJ0013-4906 &  45 &  41 & 32 & 9 &  $0.4075\pm0.0009$ & $1103\pm136$ & 1   \\
SPT-CLJ0014-4952 &  41 &  29 & 25 & 4 &  $0.7520\pm0.0009$ &  $990\pm119$ &  2  \\
SPT-CLJ0033-6326 &  37 &  18 & 13 & 5 &  $0.5990\pm0.0023$ & $1916\pm292$ & 1  \\
SPT-CLJ0037-5047 &  51 &  19 & 7 & 12 &  $1.0263\pm0.0012$ & $1350\pm182$ & 2  \\
SPT-CLJ0040-4407 &  44 &  36 & 30 & 6 &  $0.3498\pm0.0010$ & $1259\pm149$ & 1, 2  \\
SPT-CLJ0102-4603 &  44 &  20 & 15 & 5 &  $0.8405\pm0.0014$ &  $808\pm134$ & 1  \\
SPT-CLJ0106-5943 &  44 &  29 & 19 & 10 &  $0.3484\pm0.0011$ & $1297\pm197$ & 1  \\
SPT-CLJ0118-5156 &  23 &  14 & 6 & 8 &  $0.7051\pm0.0017$ &  $934\pm186$ & 1, 2  \\
SPT-CLJ0123-4821 &  29 &  20 & 16 & 4 &  $0.6550\pm0.0018$ & $1505\pm276$ & 1  \\
SPT-CLJ0142-5032 &  39 &  31 & 18 & 13 &  $0.6793\pm0.0012$ & $1000\pm110$ & 1  \\
SPT-CLJ0200-4852 &  54 &  35 & 26 & 9 &  $0.4992\pm0.0007$ & $1146\pm 97$ &  1 \\
SPT-CLJ0205-6432 &  24 &  15 & 10 & 5 &  $0.7436\pm0.0008$ &  $980\pm203$ & 1, 2 \\
SPT-CLJ0212-4657 &  36 &  26 & 14 & 12 &  $0.6535\pm0.0013$ &  $931\pm116$ & 1  \\
ACT-CLJ0215-5212 &  63 &  54 & 29 & 25 &  $0.4799\pm0.0008$ & $1043\pm 97$ &  3  \\
ACT-CLJ0232-5257 &  80 &  57 & 47 & 10 &  $0.5562\pm0.0006$ & $1020\pm 87$ &  3  \\
SPT-CLJ0233-5819 &  11 &  10 & 8 & 2 &  $0.6637\pm0.0015$ &  $754\pm183$ &  1, 2  \\
SPT-CLJ0234-5831 &  29 &  21 & 19 & 2 &  $0.4150\pm0.0009$ &  $944\pm155$ &  2  \\
ACT-CLJ0235-5121 &  98 &  78 & 71 & 7 &  $0.2776\pm0.0006$ & $1095\pm 86$ &  3  \\
ACT-CLJ0237-4939 &  68 &  62 & 55 & 7  &  $0.3343\pm0.0007$ & $1261\pm 93$ &  3  \\
SPT-CLJ0243-4833 &  43 &  39 & 26 & 13 &  $0.4984\pm0.0012$ & $1293\pm165$ & 1  \\
SPT-CLJ0243-5930 &  38 &  26 & 21 & 5 &  $0.6345\pm0.0011$ & $1155\pm135$ & 1, 2 \\
SPT-CLJ0245-5302\tablenotemark{b} &  38 &  29 & 27 & 2 &  $0.3000\pm0.0010$ & $1262\pm195$ &  1 \\
SPT-CLJ0252-4824 &  33 &  24 & 17 &  7 &  $0.4207\pm0.0006$ &  $882\pm 65$ &  1 \\
SPT-CLJ0254-5857 &  42 &  32 & 25 & 7 &  $0.4377\pm0.0015$ & $1446\pm179$ &  2  \\
SPT-CLJ0304-4401 &  46 &  35 & 32 & 3 &  $0.4584\pm0.0009$ & $1115\pm117$ &  1 \\
ACT-CLJ0304-4921 &  84 &  70 & 64 & 6 &  $0.3920\pm0.0006$ & $1051\pm 83$ &  3  \\
SPT-CLJ0307-6225 &  35 &  20 & 10 & 10 &  $0.5801\pm0.0008$ &  $652\pm153$ & 1  \\
SPT-CLJ0310-4647 &  38 &  28 & 26 & 2 &  $0.7067\pm0.0008$ &  $617\pm 72$ &  1 \\
SPT-CLJ0324-6236 &  22 &  10 & 8 & 2 &  $0.7498\pm0.0009$ & $1358\pm187$ & 1  \\
ACT-CLJ0330-5227 &  81 &  68 &  62 & 6 &  $0.4417\pm0.0008$ & $1244\pm 97$ &  3  \\
SPT-CLJ0334-4659 &  51 &  34 & 20 & 14 &  $0.4861\pm0.0014$ & $1223\pm159$ & 1  \\
ACT-CLJ0346-5438 &  92 &  86 & 66 & 20 &  $0.5298\pm0.0006$ & $1052\pm105$ &  3  \\
SPT-CLJ0348-4515 &  39 &  27 & 22 & 5 &  $0.3592\pm0.0013$ & $1246\pm167$ &  1 \\
SPT-CLJ0352-5647 &  29 &  17 & 16 & 1 &  $0.6490\pm0.0016$ &  $813\pm133$ & 1 \\
SPT-CLJ0356-5337 &  36 &   8 & 4 & 4 &  $1.0359\pm0.0042$ & $1647\pm514$ &  1 \\
SPT-CLJ0403-5719 &  43 &  29 & 21 & 8 &  $0.4670\pm0.0010$ &  $990\pm110$ &  1 \\
SPT-CLJ0406-4805 &  30 &  27 & 17 & 10 &  $0.7355\pm0.0018$ & $1216\pm135$ & 1 \\
SPT-CLJ0411-4819 &  54 &  44 & 35 & 5 &  $0.4241\pm0.0010$ & $1267\pm113$ & 1 \\
SPT-CLJ0417-4748 &  44 &  32 & 21 & 11 &  $0.5794\pm0.0012$ & $1133\pm133$ &  1 \\
SPT-CLJ0426-5455 &  17 &  11 & 7 &  4 &  $0.6420\pm0.0017$ &  $910\pm201$ &  1 \\
SPT-CLJ0438-5419\tablenotemark{b}  &  86 &  73 & 63 & 10 &  $0.4224\pm0.0007$ & $1321\pm 98$ & 1, 2, 3  \\
SPT-CLJ0456-5116 &  40 &  23 & 19 & 4  &  $0.5619\pm0.0007$ &  $822\pm149$ &  1 \\
SPT-CLJ0509-5342\tablenotemark{b} &  24 & 18 & 17 & 1 &  $0.4620\pm0.0008$ &  $690\pm 95$ &  2 \\
SPT-CLJ0511-5154 &  23 &  15 & 15 & 0 &  $0.6447\pm0.0014$ &  $758\pm137$ &  1, 2 \\
SPT-CLJ0516-5430\tablenotemark{b} & 405 & 169 & 152 & 15 &  $0.2947\pm0.0003$ &  $939\pm 56$ & 2, 4  \\
SPT-CLJ0528-5300\tablenotemark{b} &  31 &  20 & 19 & 1 &  $0.7694\pm0.0017$ & $1435\pm217$ & 2  \\
SPT-CLJ0539-5744 &  30 &  19 & 15 & 4 &  $0.7597\pm0.0021$ & $1075\pm140$ & 1 \\
SPT-CLJ0542-4100 &  38 &  31 & 25 & 6 &  $0.6399\pm0.0009$ & $1032\pm141$ & 1 \\
SPT-CLJ0549-6205 &  35 &  27 &  24 & 3  &  $0.3755\pm0.0006$ &  $666\pm104$ & 1 \\
SPT-CLJ0551-5709 &  41 &  32 & 21 & 11 &  $0.4243\pm0.0007$ &  $871\pm120$ & 2 \\
SPT-CLJ0555-6406 &  39 &  31 & 23 & 8 &  $0.3455\pm0.0008$ & $1088\pm159$ & 1  \\
SPT-CLJ0559-5249\tablenotemark{b} &  41 &  37 & 35  & 2 &  $0.6091\pm0.0011$ & $1128\pm141$ &  2 \\
ACT-CLJ0616-5227 &  20 &  16 & 13  & 3 &  $0.6835\pm0.0016$ & $1104\pm160$ &  3  \\
SPT-CLJ0655-5234 &  34 &  30 & 25 & 5  &  $0.4724\pm0.0009$ &  $883\pm103$ & 1 \\
ACT-CLJ0707-5522 &  64 &  58 & 48 & 10 &  $0.2962\pm0.0006$ &  $826\pm 79$ &  3  \\
SPT-CLJ2017-6258 &  43 &  37 & 34 & 3 &  $0.5354\pm0.0009$ &  $972\pm134$ & 1 \\
SPT-CLJ2020-6314 &  28 &  18 & 11 & 7 &  $0.5367\pm0.0013$ &  $891\pm135$ & 1 \\
SPT-CLJ2026-4513 &  25 &  19 & 14 & 5 &  $0.6887\pm0.0017$ & $1429\pm155$ & 1 \\
SPT-CLJ2030-5638 &  52 &  39 & 28 & 11 &  $0.3937\pm0.0005$ &  $620\pm 66$ & 1 \\
SPT-CLJ2035-5251 &  46 &  32 & 26 & 6 &  $0.5287\pm0.0011$ & $1015\pm111$ & 1 \\
SPT-CLJ2043-5035 &  36 &  21 & 20 & 1 &  $0.7234\pm0.0007$ &  $591\pm 71$ & 2 \\
SPT-CLJ2058-5608 &  16 &   9 & 7 & 2 &  $0.6065\pm0.0019$ & $1038\pm357$ & 1, 2 \\
SPT-CLJ2100-4548 &  41 &  20 & 14 & 6 &  $0.7122\pm0.0010$ &  $866\pm171$ & 2 \\
SPT-CLJ2104-5224 &  36 &  22 & 16 & 6 &  $0.7997\pm0.0025$ & $1244\pm172$ & 2 \\
SPT-CLJ2115-4659 &  37 &  29 & 26 & 3 &  $0.2988\pm0.0008$ &  $934\pm128$ & 1 \\
SPT-CLJ2118-5055 &  57 &  33 & 24 & 9 &  $0.6244\pm0.0010$ &  $989\pm108$ & 1, 2 \\
SPT-CLJ2136-4704 &  28 &  24 & 19 & 5 &  $0.4242\pm0.0017$ & $1448\pm224$ & 1, 2 \\
SPT-CLJ2140-5727 &  33 &  17 & 10 & 7 &  $0.4043\pm0.0012$ & $1192\pm282$ & 1 \\
SPT-CLJ2146-4846 &  29 &  26 & 23 & 3 &  $0.6230\pm0.0008$ &  $768\pm103$ & 1, 2 \\
SPT-CLJ2146-5736 &  41 &  25 & 19 & 6 &  $0.6025\pm0.0012$ &  $936\pm131$ & 1 \\
SPT-CLJ2155-6048 &  31 &  24 & 18 & 6 &  $0.5389\pm0.0011$ & $1049\pm145$ & 1, 2 \\
SPT-CLJ2159-6244 &  48 &  41 & 34 & 7 &  $0.3915\pm0.0005$ &  $794\pm 88$ & 1 \\
SPT-CLJ2218-4519 &  22 &  20 & 15 & 5 &  $0.6365\pm0.0017$ & $1172\pm143$ & 1 \\
SPT-CLJ2222-4834 &  38 &  27 & 23 & 4 &  $0.6519\pm0.0014$ & $1002\pm108$ & 1 \\
SPT-CLJ2232-5959 &  38 &  26 & 23 & 3 &  $0.5948\pm0.0010$ & $1004\pm158$ & 1 \\
SPT-CLJ2233-5339 &  42 &  31 & 25 & 6 &  $0.4398\pm0.0006$ & $1045\pm219$ & 1 \\
SPT-CLJ2245-6206 &  28 &   4 & 4 & 0  &  $0.5856\pm0.0031$ & $1363\pm953$ & 1 \\
SPT-CLJ2258-4044 &  38 &  27 & 22 & 5 &   $0.8971\pm0.0017$ & $1220\pm145$ & 1 \\
SPT-CLJ2301-4023 &  49 &  20 & 17 & 3 &  $0.8349\pm0.0023$ & $1270\pm 95$ & 1 \\
SPT-CLJ2306-6505 &  49 &  43 & 39 & 4 &  $0.5297\pm0.0008$ & $1132\pm113$ & 1 \\
SPT-CLJ2325-4111 &  48 &  33 & 26 & 7  &  $0.3570\pm0.0016$ & $1618\pm275$ & 1, 2  \\
SPT-CLJ2331-5051 &  92 &  82 & 26 &  39 &  $0.5744\pm0.0008$ & $1282\pm100$ & 2 \\
SPT-CLJ2335-4544 &  42 &  35 & 27 & 8 &  $0.5473\pm0.0010$ &  $973\pm126$ &  1 \\
SPT-CLJ2337-5942 &  80 &  39 & 35 & 4 &  $0.7768\pm0.0009$ & $839\pm105$ &  2, 4 \\
SPT-CLJ2341-5119 &  23 &  14 & 13 & 1 &  $1.0024\pm0.0012$ & $1146\pm233$ & 2 \\
SPT-CLJ2342-5411 &  16 &  11 & 5 & 6 &  $1.0758\pm0.0032$ & $1253\pm337$ & 2 \\
SPT-CLJ2344-4243 &  42 &  32 & 25 & 7 &  $0.5953\pm0.0018$ & $1814\pm219$ & 1, 2 \\
SPT-CLJ2359-5009 &  29 &  22 & 13 & 9 &  $0.7749\pm0.0011$ &  $912\pm151$ & 2 
\enddata 
\tablecomments{A summary of the results of SPT-GMOS spectroscopy by galaxy cluster. Columns from left to 
right report (1) the cluster name, (2) the total number of galaxy spectra with redshift measurements, (3) the 
number of cluster member galaxies, (4) the number of passive and post-starburst members, (5) the number 
of star-forming members, (6) the median cluster redshift, (7) the velocity dispersion estimates for 
each cluster, and (8) the reference(s) for the published galaxy spectroscopy used in our analysis.}
\tablenotetext{a}{~References: 1=\citet{bayliss16}, 2=\citet{ruel14}, 3=\citet{sifon13} 4=this work.}
\tablenotetext{b}{~Also discovered independently by ACT \citep{hasselfield2013}.}
\end{deluxetable*}

\subsection{New Magellan/IMACS Spectroscopy}

In addition to previously published galaxy spectroscopy, we also present the first publication of new spectroscopy 
in the fields of three SPT galaxy clusters. We observed SPT-CLJ0000-5748, SPT-CLJ0516-5430, and 
SPT-CLJ2337-5942 with the Inamori Magellan Areal Camera and Spectrograph \citep[IMACS;][]{dressler06} mounted 
on the Magellan-I (Baade) telescope at Las Campanas Observatory on the nights of 14-15 September 2012. 
SPT-CLJ0000-5748 and SPT-CLJ2337-5942 were each observed with one mult-slit mask, and SPT-CLJ0516-5430 
was observed with two different multi-slit masks. All masks were observed using the IMACS f/2 (short) camera and 
designed to have 1.0\arcsec\ by 6.0\arcsec\ slitlets placed using the same color-based prioritization methodology 
described in numerous previous publications \citep[Section 2.3.4 in][]{bayliss16}. 

For the two SPT-CLJ0516-5430 mask observations IMACS was configured with the WBP4791-6397 filter and 
the 300 l/mm grism, resulting in individual galaxy spectra over a wavelength range of $\Delta \lambda \sim 
4800-6400$\AA, with spectral dispersion of 1.34\AA\ per pixel, a spectral resolution element of 6.7\AA, and 
spectral resolution, $R \simeq700-1000$. SPT-CLJ0000-5748 and SPT-CLJ2337-5942 were observed 
with IMACS configured to use the WBP 6296-8053 filter and the 300 l/mm grism, resulting in individual galaxy 
spectra over a wavelength range of $\Delta \lambda \sim 6300-8100\AA$, with spectral dispersion of 1.34\AA\ 
per pixel, a spectral resolution element of 6.7\AA, and spectral resolution, $R \simeq900-1200$. 

We reduced and extracted the raw spectroscopic data using the 
COSMOS\footnote{http://code.obs.carnegiescience.edu/cosmos} package \citep{kelson03} in combination 
with custom IDL scripts based on the XIDL library\footnote{http://www.ucolick.org/\~{}xavier/IDL/}. 
The wavelength calibration was achieved using HeNeAr arc lamp exposures taken on-sky bracketing the 
science exposures, and we applied a flux calibration based on an observation of the spectrophotometric 
standard star LTT1788 \citep{hamuy1992} in the same instrument configurations as the science 
observations. Because we had a single epoch standard star observation we have only corrected for the 
wavelength-dependent response of the telescope$+$instrument system, rather than an absolute 
spectrophotometric flux calibration.

We measured redshifts for the IMACS spectra by cross-correlation with the {\it fabtemp97} and 
{\it habtemp90} templates in the RVSAO package for IRAF \citep{kurtz98}. All redshifts were corroborated 
by visually confirming the presence of strong features by eye. Following previous work we note that the 
redshift uncertainties output by RVSAO underestimate the true statistical uncertainty by a factor of 
$\sim$1.7 \citep{quintana00}, and so we have corrected the RVSAO outputs by multiplying the uncertainties 
by this factor. We show a representative portion of the catalog measurements made from these 
new IMACS spectra in Table~\ref{tab:catalog}, and the full catalog of new IMACS data will be publicly released 
in parallel with this paper. From these IMACS data we have added 474 new galaxy spectra, 172 of which 
are cluster members. The inclusion of the new IMACS spectra described above and these literature 
data have grown the number of cluster member spectra in our raw sample by $\sim$80\%. The full cluster dataset 
is described in Table~\ref{tab:spectable}, and all 89 clusters are plotted as a function of SZ mass and redshift in 
Figure~\ref{fig:mzdist}. The full SPT-GMOS$+$literature sample consists of 4148 galaxies, 2868 of 
which are cluster members.

\subsection{Selecting Cluster Member Galaxies}

Previous studies have rightly noted that member selection is a critical step in assembling large samples 
of cluster galaxies with which to test for velocity segregation effects 
\citep{biviano06a,biviano09,saro13,girardi15,barsanti16}. Member selection must be treated 
thoughtfully in order to minimize potential biases in the dispersion estimates that could arise by including, 
for example, large numbers of interloper galaxies. We apply a two-stage member selection process designed 
to emulate the ``peak$+$gap'' procedure that has 
been used in many previous studies \citep{fadda96,biviano13,girardi15,barsanti16}. This procedure relies on 
a first stage selection that takes place solely in line-of-sight velocity space, identifying velocity 
distributions associated with clusters as over-densities in the available radial velocity catalogs. We 
iteratively compute the bi-weight estimates of the center and width (i.e., median redshift and velocity 
dispersion) of the velocity distribution associated with each cluster in the same way as described in 
\citet{bayliss16}. Briefly, we iteratively applied the bi-weight location and scale estimators from 
\citet{beers90}, rejecting $>3\sigma$ outliers until converging on a stable velocity dispersion.

\begin{figure*}[t]
\centering
\includegraphics[scale=0.62]{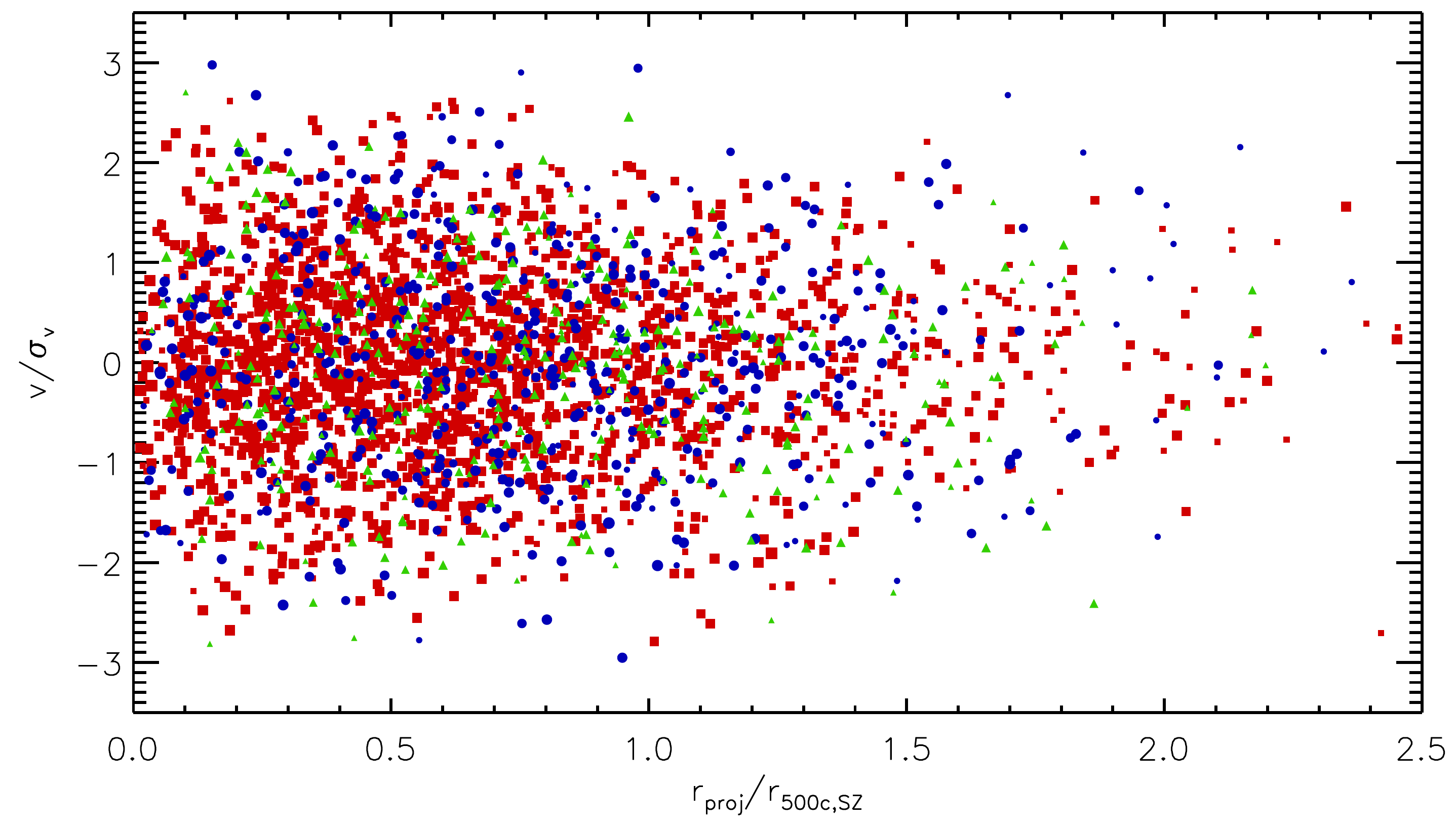}
\caption{\scriptsize{
The full ensemble of member galaxies from all clusters used in this analysis with each galaxy 
plotted according to its peculiar velocity normalized to the velocity dispersion of its host cluster, 
and its projected radial separation relative to $r_{500c,\mathrm{SZ}}$ for its host cluster. Individual galaxies 
are plotted in one of three colors indicating whether they were classified as passive (red squares), 
post-starburst (green triangles), or actively star-forming (blue circles), and the size of each plotted point 
is proportional to the brightness of the galaxy, with brighter galaxies having larger plot symbols. Galaxies 
that lack good photometry are plotted as the smallest points.}}
\label{fig:phasespace}
\end{figure*}

We then take the results of the first stage selection as a starting point and apply the ``shifting gapper'' 
procedure to select member galaxies for each cluster. The shifting gapper uses both radial 
velocity and projected radial information to iteratively include or reject galaxies as cluster members until 
convergence is reached. The shifting gapper is applied by iteratively testing each candidate member galaxy 
using other candidate member galaxies that are nearby neighbors in projected radius, flagging the tested 
galaxy as a cluster member if the largest gap in peculiar velocity among the neighbor galaxies does not 
exceed some fixed value. Specifically, we apply the same implementation of the shifting gapper technique 
as described by \citet{crawford14}, which differs slightly from the original \citet{fadda96} implementation in 
that it uses absolute peculiar velocities rather than the raw velocity values to reflect the fact that on average, 
galaxy clusters are radially symmetric in velocity space \citep{kasun05}. Our implementation also uses running 
bins of the nearest $n$ galaxies in radius rather than rigid radial bins (e.g., fixed $\Delta r =0.4$ or 0.5 Mpc 
h$^{-1}$) to avoid the effects of artificial discontinuities across bins in sparsely sampled clusters, which most 
of our clusters are. For the gap velocity criteria we use 1000 km s$^{-1}$, the value commonly applied in the 
literature \citep{fadda96,crawford14,barsanti16}. We 
explore different values of $n$ in our clusters with $N>30$ spectroscopic galaxies, and find the result to be 
insensitive to the value of the number of neighbors used from $n = 10-30$. Henceforth we 
use $n = 10$ because it allows us to maximize the number of galaxies used in our analysis by including 
all 84 galaxy clusters with $N>10$ member galaxies.

The final velocity dispersion estimates for two clusters---SPT-CLJ0033-6236 and SPT-CLJ2344-4243---at 
$1916\pm292$ km s$^{-1}$ and $1814\pm219$ km s$^{-1}$, respectively, are surprisingly high relative 
to what we would normally expect for typical SPT galaxy clusters. We have considered these estimates 
carefully and we do not believe that it is problematic to end up with dispersion estimates for two out of 84 
clusters scattering high as these do. This belief is based in part on the fact that our dispersion estimates 
are generally based on relatively small numbers of member galaxies, and that the 90\% confidence intervals, 
for instance, of both dispersion estimates, extend as low as $\sim$1300-1400 km s$^{-1}$. Furthermore, 
one of the high dispersion clusters, SPT-CLJ2344-4243, is among the most massive clusters in the SPT 
sample and an extremely large dispersion estimate is not necessarily surprising in its case.

\subsection{Galaxy Spectral Type Classification}
\label{sec:classification}

We use the available 1D spectra for all galaxies included in this analysis to measure the strength of standard 
spectral features for the complete SPT-GMOS$+$literature sample. We 
apply the same spectral index analysis as described in \citet{bayliss16} to measure the rest-frame 
equivalent width, $W_{0}$, of two important features: the [O {\small II}] $\lambda \lambda$ 
3727,3729 doublet and the H-$\delta$ n $= 6 \longleftrightarrow 2$ hydrogen transition. These two 
transitions are well-studied \citep{balogh99} and can be used to uniquely assign galaxies one of six different 
types based on the criteria first proposed by \citet{dressler99}. We use the 
\citet{dressler99} criteria to classify each galaxy in our sample as either k, k$+$a, a$+$k, e(c), e(b), or 
e(a), where k-type indicates a passive galaxy, k$+$a- and a$+$k-type indicate a post-starburst galaxy, and 
each of e(c)-, e(b)-, and e(a)-type indicate a star-forming galaxy. The specific criteria from 
\citet{bayliss16} are reproduced in Table~\ref{tab:galtypes}, in which an entry of ``none'' 
indicates no \oii\ emission feature at $< 2 \sigma$ significance,``yes'' indicates a detection of \oii\ at 
a $\geq2\sigma$ significance, and ``any'' simply indicates that any value for the \hdelta\ feature is acceptable 
for the e(b) spectral type.

These classifications are similar to color-based identifications that use the location of galaxies in color 
magnitude space, but are instead based on the strength of two distinct features that are associated 
with the presence of O, B, and A stars. In contrast to the simple ``red'' or ``blue'' color-based classification 
we use the strength of well-studied spectral features to accurately identify actively star-forming galaxies, 
passive galaxies, and the population of galaxies that is thought to be transitioning between the two 
(post-starburst galaxies), based on the presence (or absence) of young and intermediate age stars. Some 
studies have also preferred to refer to different galaxy types using the morphological late-type vs. early-type 
classification \citep{hubble26}, where late-type galaxies tend to be those we label as star-forming and 
early-type generally correspond to those we call passive, with the post-starburst galaxies approximately 
occupying the ambiguous transition space in between as S0 galaxies \citep{larson80,bothun82,muzzin14}. 
Henceforth in this paper we will refer to galaxies as passive, post-starburst, and star-forming based on 
well-defined measurements of spectral indices (Table~\ref{tab:galtypes}). In addition to exploring the peculiar 
velocities for galaxies of each of these three types, we will also perform some analyses where we explore the 
effects of splitting the sample into ``red'' (passive) vs. ``nonred'' (post-starburst plus star-forming) 
subsamples, as well as ``blue'' (star-forming) vs. ``nonblue'' (post-starburst and passive) subsamples 
in order to compare our results to previous color- and morphology-based work and simulations.

Wherever possible we also include brightness measurements for galaxies in the spectroscopic sample, 
following the same methodology described in \citet{bayliss16}. Briefly, we use photometry in the $r-$ and 
$i-$ bandpasses of the Sloan Digital Sky Survey \citep[SDSS;][]{york2000} photometric system, where 
many clusters have imaging from one or more telescopes/instruments \citep{song12,sifon13,bleem15} 
that was taken using filters that are similar to---and were calibrated against---the SDSS $r$ and $i$ bandpasses. 
For some SPT clusters we have imaging in the Cousins/Johnson $BVRI$ system, which we convert into the SDSS 
system using empirically determined formulas from \citet{jordi2006}. Our final cluster sample has photometric 
data that varies significantly in its relative depth and seeing, which 
results in a highly non-uniform photometric completeness. Because we focus here on measuring relative 
brightnesses between cluster member galaxies this limitation should not strongly affect our analysis, 
but is worth bearing in mind for anyone using this catalog for other analyses.

In total we have magnitude measurements 
in at least one of the $r-$ or $i-$ bandpasses for $\sim$76\% of the galaxies in our spectroscopic sample; most of 
the remaining galaxies appear in cluster fields where we do not have sufficiently deep or well-calibrated photometric 
catalogs to recover an $r-$ or $i-$band equivalent magnitude. A few galaxies lack brightness measurements because 
they are significantly fainter than the galaxies that were typically targeted on spectroscopic masks and do not appear 
in the available photometric catalogs; these cases consist primarily of galaxies that serendipitously fell into slits that 
were targeting other brighter galaxies, resulting in a spectrum and redshift measurement. 

\begin{figure*}[t]
\centering
\includegraphics[scale=0.55]{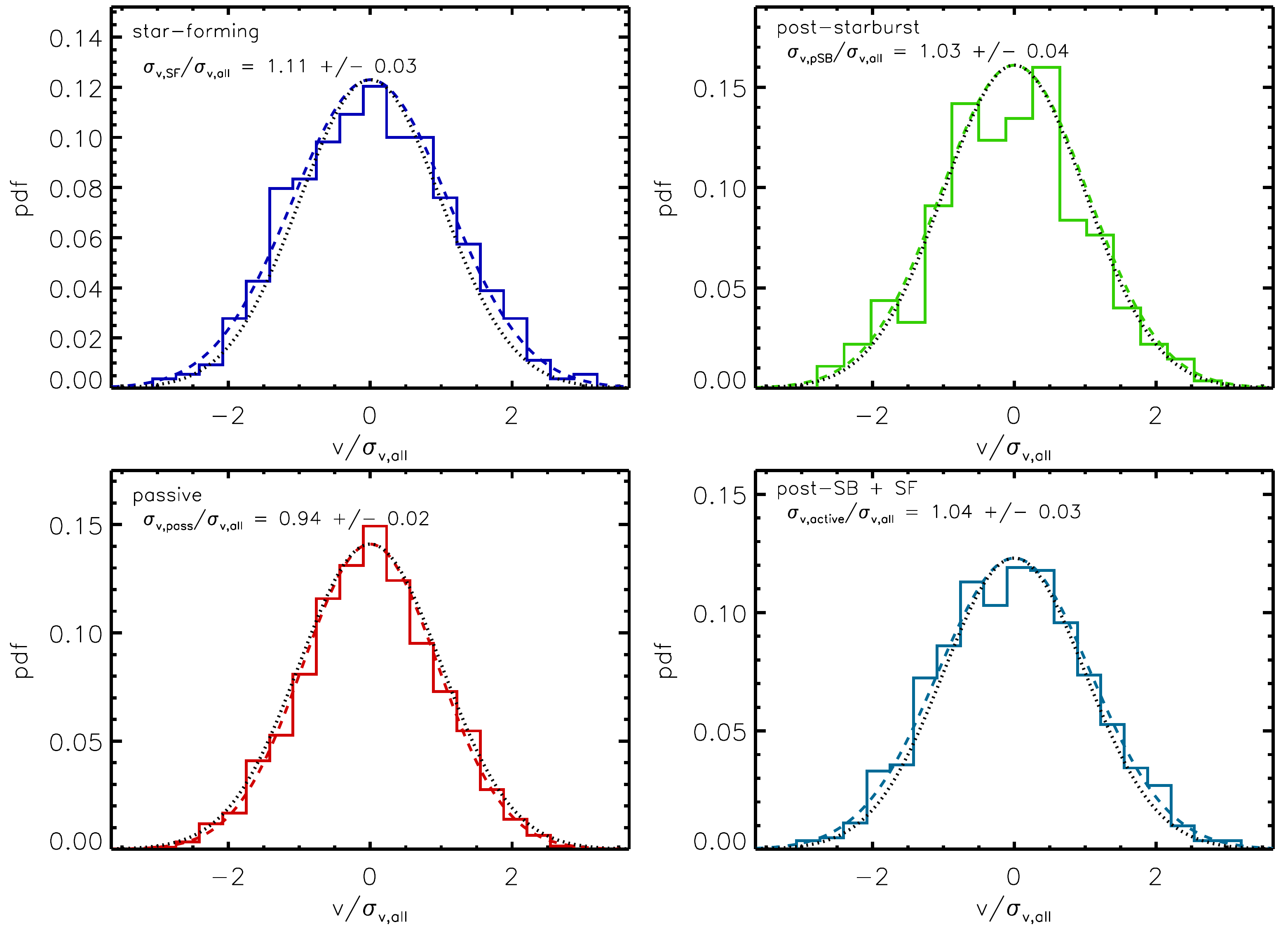}
\caption{\scriptsize{
Stacked velocity distributions for cluster member galaxies separated by spectral type. Each panel includes a 
colored dashed line indicating a Gaussian distribution with the dispersion estimated from the data (solid 
histograms), as well as a black dotted line that indicates the fiducial velocity distribution, $\sigma_\mathrm{v,all}$, 
of all cluster member galaxies include in our ensemble. The ratio of the velocity dispersion of 
each subset of member galaxies to the dispersion estimated from the stack of all 
cluster member galaxies is also inset. 
\emph{Top Left:} The stacked velocity distribution of star-forming cluster members.
\emph{Top Right:} The stacked velocity distribution of post-starburst cluster members.
\emph{Bottom Left}: The stacked velocity distribution of passive cluster members.
\emph{Bottom Right:} The stacked velocity distribution of post-starburst $and$ star-forming cluster 
members, i.e., all non-passive members.}}
\label{fig:colorstacks}
\end{figure*}

\section{Velocity Segregation Effects}
\label{sec:vse}

\subsection{Creating Ensemble Galaxy Cluster Samples}

Previous studies have had more success detecting velocity segregation when using large samples of galaxy 
clusters stacked to form ensemble cluster velocity distributions and phase-spaces. 
There is a large \citep[$\sim$30-40\%;][]{white10,saro13,gifford13a,ruel14} systematic uncertainty in 
individual cluster velocity dispersions that predominantly results from the uncertainty associated with 
measuring line-of-sight recessional velocity for systems that are triaxial in nature. The orientation of 
an individual cluster relative to the angle normal to the plane of the sky will skew the measured velocity 
dispersion \citep{noh12}. If knowledge of the orientation of individual clusters can somehow be inferred 
\citep[using gravitational lensing information, for example;][]{bayliss11b} then it may be possible to mitigate 
this intrinsic scatter in line-of-sight velocity dispersion estimates, as the inertia and velocity tensor are quite 
well aligned \citep{kasun05}. However, for most galaxy cluster samples there is no such information available 
and the cluster-to-cluster noise can very easily drown out velocity segregation effects in individual systems. 
This scatter can even systematically suppress or amplify (depending on the exact orientation) the signatures 
of velocity segregation in individual clusters. 

\begin{deluxetable}{cccc}
\tablecaption{Galaxy Spectral Type Classification\label{tab:galtypes}}
\tablewidth{0pt}
\tabletypesize{\small}
\tablehead{
\colhead{Spectral} &
\colhead{O[{\small II}] W$_{0}$ } &
\colhead{H-$\delta$ W$_{0}$ } &
\colhead{Classification} \\
\colhead{Type} &
\colhead{(\AA)} &
\colhead{(\AA)} &
\colhead{ } }
\startdata
k          & none  &  $<$ 3  & passive \\
k$+$a &  none  &   $\geq$ 3, $\leq$ 8  & post-starburst \\
a$+$k &  none  &  $>$ 8  & post-starburst \\
e(c)     &  $>$ -40  & $<$ 4  & star-forming \\
e(b)     &  $\leq$ -40  & any & star-forming \\
e(a)     & yes  & $\geq$ 4  & ~star-forming 
\enddata
\tablecomments{This table is reproduced from \citet{bayliss16} and lists the criteria we use to assign galaxy 
types. From left to right the columns are: (1) the specific galaxy spectral type, (2) [O II]$\lambda$ 3727 equivalent 
width criterion, (3) the \hdelta\ equivalent width criterion, and (4) the broad galaxy classification (i.e., passive, 
post-starburst, or star-forming).}
\end{deluxetable}

Our goal is to leverage the statistical power of our full sample of galaxy cluster spectroscopy and 
test for the presence of velocity segregation effects that are expected to be significantly smaller 
than the uncertainties in velocity dispersion measurements of individual galaxy clusters. Constraining 
these average behaviors could provide valuable information, for instance, for investigating systematic effects 
on velocity dispersion measurements for cosmological samples of galaxy clusters, where the objective is 
to calibrate average cluster observable measurements against simulations \citep{borgani97}. 

Combining clusters into ensembles naturally averages out individual cluster-to-cluster behavior. This 
averaging-out effect could be considered detrimental for some purposes, but is ideal for analyses 
such as ours where the goal is precisely to understand the {\it average} behaviors of cluster member 
galaxies. It is therefore not surprising that the majority of detections of velocity segregation effects in 
the literature result from analyses of ensemble clusters obtained by stacking the velocity distributions 
of many clusters. 

Studies that analyze well-sampled individual galaxy clusters (i.e., having 
$\gtrsim$100 member spectra) find mixed results with respect to measuring velocity segregation 
\citep{zabludoff93,hwang08,owers11,girardi15}, which makes sense in the context of a physical 
picture in which random orientation effects could suppress or enhance the velocity segregation signal 
for individual clusters. Consider, for example, the prevailing understanding in which passive and 
star-forming galaxies have different velocity profiles, reflecting the idea that passive galaxies are 
spatially distributed so as to more closely follow the shape of the underlying gravitational potential, 
while star-forming galaxies are distributed more in the outskirts, likely include a population of actively 
in-falling galaxies, and are therefore less reflective of the cluster potential. The passive galaxy population 
in such a cluster observed with the long axis oriented normal to the plane of the sky (i.e., a radial football) 
would have its 
measured velocity dispersion boosted upward more than the dispersion of the cluster's star-forming galaxy 
population, an effect that would cancel out the expected velocity segregation signal.

We create ensemble velocity distributions and phase-spaces by normalizing the peculiar velocity, $v_{p}$, 
of each cluster member galaxy by the estimated velocity dispersion of its host cluster and the projected 
radial separation, $r_\mathrm{proj}$, between each galaxy and SZ-derived centroid of its host cluster, 
normalized by $r_{500c,\mathrm{SZ}}$ for its host cluster. We compute $r_{500c,\mathrm{SZ}}$ for each 
cluster as the radius within which the mean density is equal to 500 times the critical density at the cluster redshift, 
based on the SZ mass estimate for each cluster \citep{hasselfield2013,bleem15}. This produces normalized 
$v_{p}$ and $r_\mathrm{proj}$ values for each galaxy that are derived from independent 
observations: velocity dispersions and the SZ, respectively. Our cluster sample is comprised entirely 
of SZ-selected systems that are among the most massive clusters in the Universe \citep[M$_{500c} 
\gtrsim 3 \times 10^{14}$ \msun~h$_{70}^{-1}$;][]{bleem15}. This commonality in selection and mass 
supports the assumption that the ensemble phase-spaces that we construct are reliable representations 
of the average properties of the clusters used to populate them. The full-sample ensemble phase-space 
for our cluster sample is shown in Figure~\ref{fig:phasespace}.

\subsection{Velocity Segregation By Spectral Type}
\label{sec:vs_by_type}

\begin{deluxetable}{cccc}
\tablecaption{Ensemble Dispersions By Galaxy Sub-population\label{tab:stats}}
\tablewidth{0pt}
\tabletypesize{\small}
\tablehead{
\colhead{Galaxy Subset} &
\colhead{N$_{gals}$ } &
\colhead{ $\sigma_{v}/\sigma_\mathrm{v,all}$} &
\colhead{AD Prob\tablenotemark{a} } }
\startdata
Passive & 2028  & $0.947\pm0.015$  &  0.385   \\
Post-starburst &  274  & $1.034\pm0.044$ & 0.25 \\
Star-forming &  539  &  $1.110\pm0.031$ & 0.388 \\
{\scriptsize Bright  ($m < m^{*}-0.5$)} & 298 & $0.894\pm0.041$ & $<1$e$-9$\tablenotemark{b} \\
{\scriptsize Faint ($m > m^{*}+0.5$)} & 1107 & $1.01\pm0.021$ & 0.33 \\
\enddata
\tablecomments{Here we show the results of statistical tests of different sub-populations of galaxies in 
the ensemble stack. From left to right the columns are: (1) the sub-population of galaxies analyzed, (2) 
the total number of galaxies of that sub-population that are in the ensemble, (3) the velocity dispersion 
of the galaxies in that sub-population, normalized to the velocity dispersion of the full ensemble, and 
(4) the probability that galaxies in that sub-population are drawn from a Gaussian velocity distribution 
based on the Anderson-Darling (AD) statistic.}
\tablenotetext{a}{A value of 0.1 indicates a 10\% probability that galaxies in a given sub-population 
have a peculiar velocity distribution that is consistent with being Gaussian.}
\tablenotetext{b}{The approximation that we use from \citet{hou2009} becomes imprecise at 
very large values of statistical significance.}
\end{deluxetable}

\begin{figure}[h]
\centering
\includegraphics[scale=0.473]{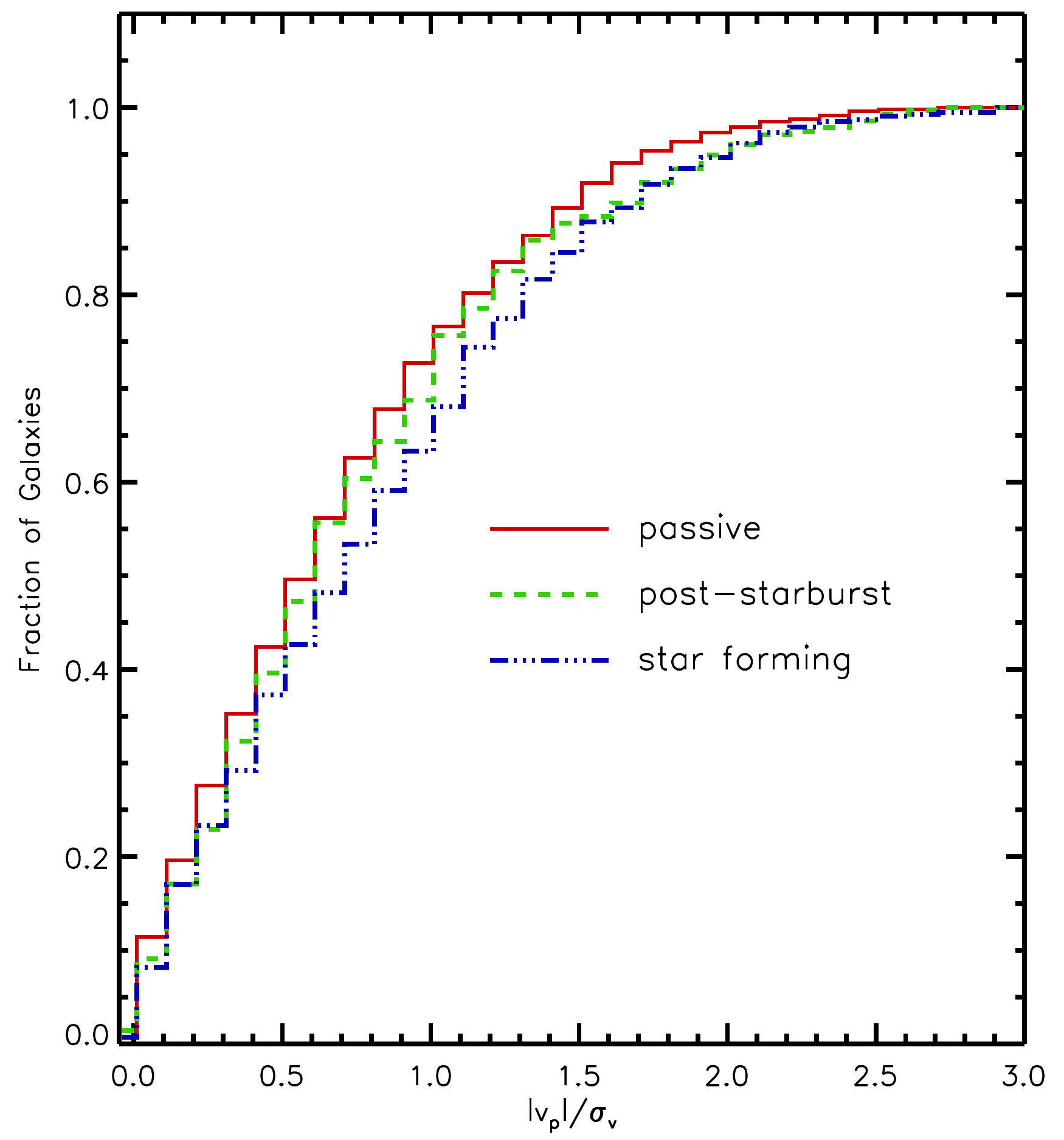}
\caption{\scriptsize{
The cumulative distribution functions of the absolute peculiar velocities for cluster member galaxies, where 
galaxies have been separated into passive, post-starburst, and star-forming sub-samples.}}
\label{fig:cdf}
\end{figure}

The ensemble cluster contains 2847 galaxies, 2841 of which have spectral classifications. We analyze 
the ensemble of 2841 galaxies with known spectral types, and generate the velocity distributions for 
passive, post-starburst, and star-forming cluster members, shown in Figure~\ref{fig:colorstacks}. 
We also plot the cumulative distribution functions (CDFs) of the absolute peculiar velocities of the same three 
types of cluster member galaxies (Figure~\ref{fig:cdf}). Qualitatively there is a clear trend in both 
Figures~\ref{fig:colorstacks} \& \ref{fig:cdf}, with the star-forming galaxies having a larger velocity dispersion 
than the full ensemble, and tending to have larger absolute peculiar velocities than post-starburst and 
passive galaxies. Passive galaxies, on the other hand, have a smaller velocity dispersion relative to the 
full ensemble and tend to have smaller absolute peculiar velocities. Post-starburst galaxies represent 
the evolutionary ``in-between'' step between star-forming and passive galaxies and, interestingly, they also 
tend to lie in between passive and star-forming galaxies in velocity space for massive galaxy clusters. 

Quantitatively we measure the magnitude of velocity segregation by galaxy type by performing a resampling 
of the ensemble distribution. Specifically, we generate 1000 Monte Carlo realizations of velocity distributions 
using only galaxies of the same type, where each realization is constructed by drawing 125 galaxies {\it without 
replacement} from the ensemble cluster, where ``without replacement'' here simply means disallowing any 
galaxy in the ensemble dataset from appearing more than once in a resampled realization. It is important 
to draw without replacement when performing a resampling analysis in which the 2nd order statistic (i.e., 
the dispersion) is being investigated, because allowing replacement draws will artificially bias the recovered 
dispersions low. The choice to use 125 galaxies is made to ensure good statistical precision in the dispersion 
estimate from each resampled dataset, while also allowing us to generate sufficient unique realizations from 
our ensemble so as to recover a good estimate of the spread in dispersion values---i.e., the statistical 
uncertainty in the velocity dispersion for each sub-population of galaxies.

We find that passive cluster members have a normalized velocity dispersion of $0.947$ $\pm$ $0.015$ 
relative to the entire cluster sample, while the corresponding velocity dispersion for star-forming galaxies 
is $1.110$ $\pm$ $0.031$. Post-starburst galaxies actually have a velocity distribution that is indistinguishable 
from the total member galaxy distribution ($\sigma_{v,PSB}/\sigma_{all} = 1.034$ $\pm$ $0.044$), which is 
likely the result of a combination of two factors: 1) the post-starburst measurement is noisier because 
post-starbursts are $<10\%$ of our sample, and 2) post-starburst galaxies are a transitional population 
between the star-forming and passive populations. The differences 
between the passive and star-forming galaxy velocity dispersion is significantly larger than the statistical 
uncertainties ($>3\sigma$). Directly comparing the velocity dispersion of passive and star-forming galaxies 
we find the star-forming galaxies have a dispersion $17$ $\pm$ $4$\% larger than the passive cluster galaxy 
population.

One other factor to consider is the target prioritization strategy we used when designing the multi-slit masks that 
produced our spectra. As described in \citet{ruel14} and \citet{bayliss16}, we selected targets based on photometric 
colors. Red-sequence selected candidate member galaxies were given top priority, with potential blue-cloud 
galaxies making up the next priority level, and all other sources used as filler objects. In principle this strategy could 
bias our radial sampling to preferentially yield spectra of passive (i.e., red-sequence) galaxies at smaller projected 
cluster-centric radii. However, in practice we find that the projected radial distribution of slits with different priorities 
is relatively flat. We also note that the average peculiar velocities of all cluster member galaxies, regardless of type, 
tend to decrease with increasing projected radius, so that the impact of a prioritization bias that skewed our non-passive 
galaxy samples to larger projected radii would only serve to slightly suppress any velocity segregation signal.

It is also important here to consider the possible impact of interloper galaxies that 
we have mistakenly included as members in our galaxy spectroscopy sample. We cannot realistically expect to 
identify member galaxies with perfect accuracy, and so interlopers are almost certainly present at some level. 
That being said, we also point out that analysis of the effects of interlopers in simulations suggests that they do 
not strongly bias velocity dispersion estimates with radial sampling similar to ours. \citet{saro13} studied the 
effects of red sequence (i.e., passive galaxy) interlopers in simulations, finding that interloper fractions remain 
small ($\lesssim$7\%) within projected radii $r_\mathrm{proj} \lesssim r_{200c}$ (where $r_{200c}$ is the radius within 
which the mean density is 200 times the critical density); this suggests that we can expect our cluster data to 
typically include of order 1-2 interlopers per cluster. \citet{saro13} found that these small numbers of 
red-sequence interlopers, when included, act to boost the estimated velocity dispersions of clusters. 
\citet{biviano06a} explored the effects of blue/star-forming galaxies, which are much more common in the field, 
as an interloper population and found that they actually have the effect of suppressing the estimated velocity 
dispersions of clusters. These simulation results indicate that the effects of passive and star-forming interloper 
galaxies may both act to suppress an underlying velocity segregation signal by galaxy type.

Statistical tests are also helpful to quantify the significance of our detection of velocity segregation by galaxy type, 
and here we employ both the Kolmogorov-Smirnov (KS) and the Anderson-Darling (AD) tests. The KS test is a tool 
that is commonly used to quantitatively compare distributions. It is useful for identifying differences near the centers 
of probability distributions, but lacks power in the wings. The AD test, on the other hand, is sensitive to deviations 
from the centers out through the wings of distributions, and has been demonstrated to be a much more powerful 
test for detecting differences 
between CDFs than the KS test \citep[e.g.,][]{hou2009}. Both the KS and AD test results are consistent with each 
of the passive, post-starburst, and star-forming galaxy populations having Gaussian velocity distributions. Of these 
the post-starburst galaxies have the largest probability of rejecting the null hypothesis at 25\% ($\sim1.2\sigma$). 
The normalized velocity dispersions for passive/post-starburst/star-forming galaxies are given in Table~\ref{tab:stats}, 
along with the AD test probabilities that each galaxy type's velocity distribution is Gaussian.

When we compare the distributions of different galaxy types against one another we find that the KS test identifies 
a $\sim$1.6$\sigma$ difference between the passive and star-forming galaxy distributions, while the comparison 
with the AD test produces a $\gtrsim6\sigma$ difference, but neither the KS nor AD tests identify statistically 
significant differences between the other pairings of galaxy subsets. We point 
out here that the formula that we use to convert AD statistic values into confidence levels for rejecting the 
null hypothesis \citep{hou2009} is a numerical approximation that becomes imprecise once the statistical 
significance becomes very large, but is more than adequate for our purposes where we are primarily interested 
in testing whether we can confidently reject the null hypothesis.

\begin{figure}[h]
\centering
\includegraphics[scale=0.44]{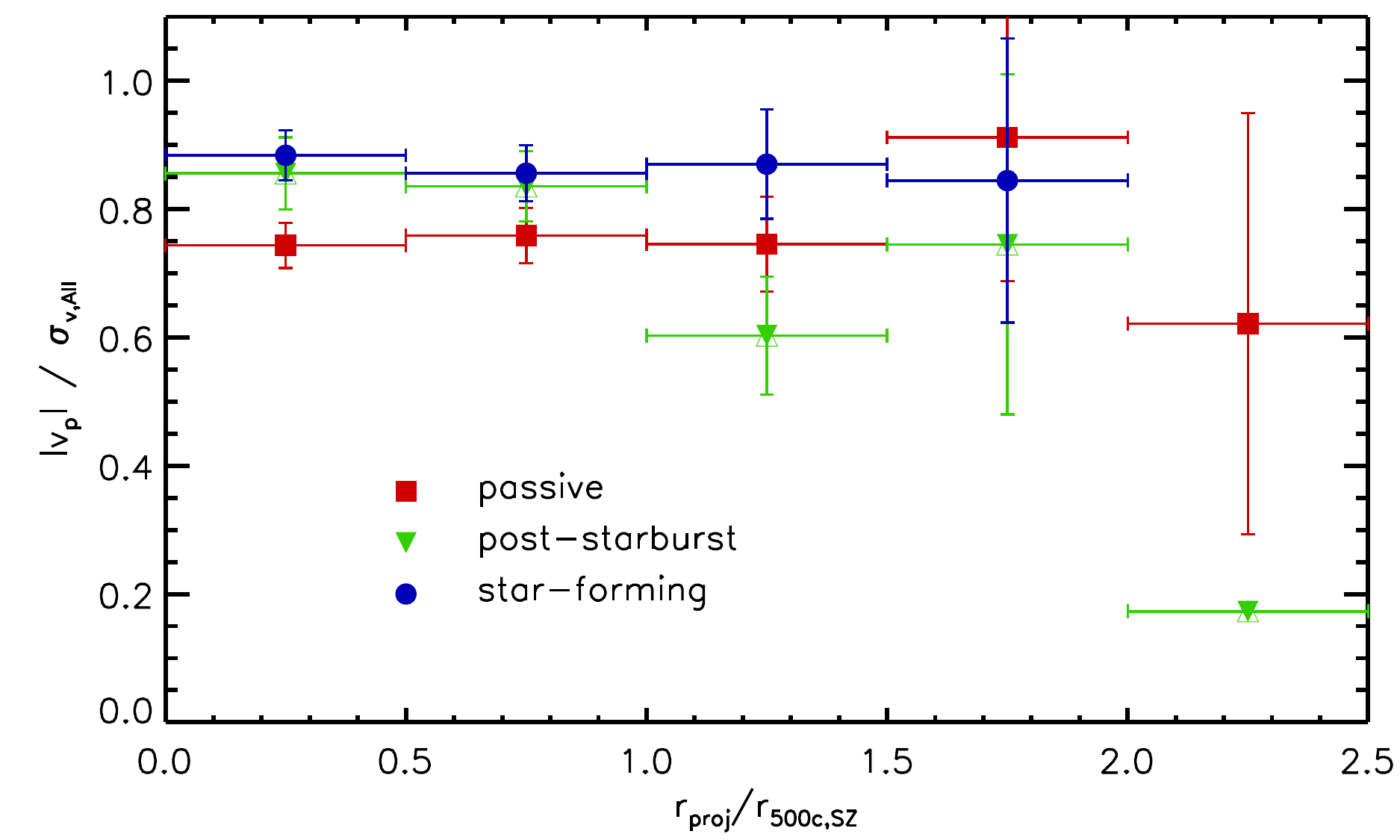}
\caption{\scriptsize{
The average absolute peculiar velocities of passive/post-starburst/star-forming galaxies in the ensemble 
phase-space where galaxies have been sorted into bins of radius relative to $r_{500c,\mathrm{SZ}}$. 
The expectation value for the peculiar velocity of galaxies of each type is computed as the 
mean of all galaxies of a given type within a radial bin, with uncertainties estimated from bootstrapped 
realizations of each galaxy sub-population in each bin.}}
\label{fig:radbins}
\end{figure}

We also use the AD test to quantify deviations from Gaussianity in the ensemble velocity distributions of 
passive, post-starburst, and star-forming galaxies, and find that all three spectral types are consistent with 
Gaussian velocity distributions. 

\begin{figure}
\centering
\includegraphics[scale=0.45]{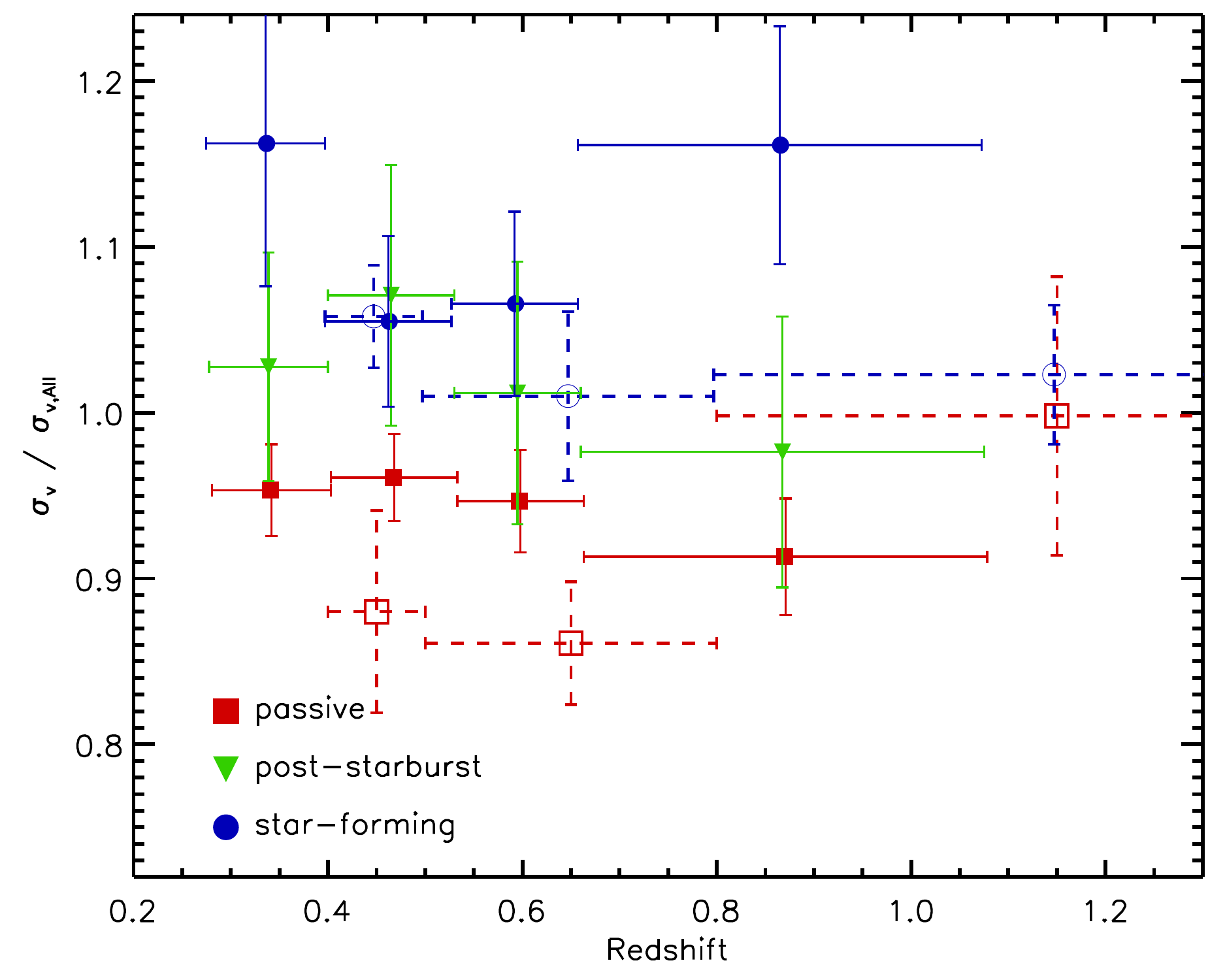}
\caption{\scriptsize{
Velocity dispersion estimates, normalized to the full ensemble, for cluster member galaxies grouped by 
spectral type in four different redshift bins. Our data are plotted as filled symbols, while we over-plot 
as open symbols (with dashed error bars) equivalent measurements of normalized red and blue galaxy 
dispersions in three redshift bins from \citet{barsanti16}. Our data show a clear separation between the 
velocity distributions of passive and star-forming galaxies out to $z\sim1$.}}
\label{fig:zbins}
\end{figure}

\subsubsection{Trends With Projected Cluster Radius}

Having detected segregation effects between velocity distributions of galaxies differentiated by spectral type, 
we now expand our analysis to test for possible trends as a function of other 
quantities. The average peculiar velocities of passive, post-starburst, and star-forming galaxies are plotted in 
radial bins of width $0.5r_{500c,\mathrm{SZ}}$ in Figure~\ref{fig:radbins}. From Figure~\ref{fig:radbins} we 
see that the passive galaxies have consistently smaller dispersions than the star-forming galaxies in the more 
central radial bins where our spectroscopic data is more numerous and the uncertainties smallest 
($r_\mathrm{proj} \leq 1.5 r_{500c,\mathrm{SZ}}$). The outer radial bins are less well sampled, with larger 
uncertainties, and so it is difficult to draw firm conclusions from the data, and are also where the sample 
should be more likely to suffer from some contamination by interloping galaxies 
because the cluster over-density is smallest at larger radii. While we see no strong trends in radius for any 
sub-population of galaxies, we do note a low-significance drop in the peculiar velocities of star-forming galaxies 
at larger radii, while passive galaxy peculiar velocities are flat with radius, within the uncertainties. This trend 
would be consistent with studies that find star-forming 
galaxies to preferentially have radially-elongated orbits, while passive galaxy orbits have larger tangential 
components \citep{katgert96,biviano04}, and makes sense in the context of a physical picture in which 
star-forming cluster member galaxies reflect a population of galaxies that have recently fallen into the 
cluster potential while passive galaxies tend to have been in the cluster for longer times and are more 
dynamically relaxed \citep{mahajan11,haines15}. 

\subsubsection{Trends With Cluster Redshift}
\label{sec:zbins}

Our galaxy cluster sample has the benefit of an approximately flat selection in mass, despite spanning a relatively 
wide range in redshift. We can, therefore, divide the sample up into redshift bins to investigate possible trends in 
the velocity segregation effects in clusters of similar masses as a function of redshift. We define four redshift bins 
that are chosen to place a similar number of cluster member galaxies in each bin; the redshift intervals we use 
and the velocity dispersion of galaxies of each spectral type are given in Table~\ref{tab:zbins}. Each bin contains at 
least 83 galaxies of each spectral type, ensuring good statistical significance on the estimated dispersion of each 
sub-population in each redshift bin.

The data are shown in Figure~\ref{fig:zbins}, and it is clear that the tendency for star-forming galaxies to have 
larger velocity distributions than passive galaxies holds out to $z\gtrsim1$ in our data, though the individual 
estimates become noisier due to poorer statistical sampling as a result of dividing the sample. For comparison 
we also show the recently published results from \citet{barsanti16} as open symbols with dashed error bars, 
which are generally in good agreement with 
our measurements, especially at $z\lesssim0.8$. The offsets between our data and the \citet{barsanti16} 
measurements are likely a  result of slightly different compositions of our respective datasets with respect to the 
fraction of the data for passive vs. star-forming member galaxies. That is to say, both analyses use all 
available member galaxies to compute $\sigma_\mathrm{v,all}$ for each cluster, which normalizes the peculiar 
velocities in the ensemble, and if the number of passive vs. star-forming galaxies that are used is different 
between our two datasets then the fiducial $\sigma_\mathrm{v,all}$ velocity dispersion will change. A sample 
with more blue/star-forming galaxies would have, on average, a systematically larger $\sigma_\mathrm{v,all}$ 
estimated for each cluster, while a more red/passive galaxy heavy dataset would have smaller values of 
$\sigma_\mathrm{v,all}$. Larger values of $\sigma_\mathrm{v,all}$ would obviously result in smaller values 
of the ratios of the dispersions of both passive and star-forming subsets, relative to $\sigma_\mathrm{v,all}$, 
while the opposite is true for smaller values of $\sigma_\mathrm{v,all}$. Notably, the \citet{barsanti16} sample 
tends to include relatively more blue/star-forming member galaxies than ours, and accordingly their 
measurements of the red/blue galaxy dispersions normalized to $\sigma_\mathrm{v,all}$ are shifted 
systematically low relative to ours. Ideally we might prefer to use the dispersions measured using only passive 
galaxies as the normalizing dispersion for each individual cluster. However, doing so with our current dataset 
would result in much higher statistical uncertainty in the normalizing dispersion used for each cluster, and 
severely degrade our ability to detect the signatures of velocity segregation.

\begin{deluxetable*}{cccccccc}
\tablecaption{Ensemble Galaxy Clusters Grouped By Redshift\label{tab:zbins}}
\tablewidth{0pt}
\tabletypesize{\small}
\tablehead{
\colhead{Redshift Interval} &
\colhead{N$_{gals}$ } &
\colhead{N$_{PASS}$ } &
\colhead{N$_{PSB}$ } &
\colhead{N$_{SF}$ } &
\colhead{ $\sigma_{v,PASS}/\sigma_\mathrm{v,all}$} &
\colhead{ $\sigma_{v,PSB}/\sigma_\mathrm{v,all}$} &
\colhead{ $\sigma_{v,SF}/\sigma_\mathrm{v,all}$} } 
\startdata
$z \leq 0.4$               & 757 & 599 & 85 &  93 &  $0.953\pm0.028$ &  $1.027\pm0.069$ & $1.162\pm0.086$  \\
$ 0.4 < z \leq 0.53 $  & 842 & 604 & 87 & 168 & $0.961\pm0.026$ &  $1.071\pm0.078$ & $1.055\pm0.052$ \\
$ 0.53 < z \leq 0.66$ & 718 & 483 & 86 & 165 & $0.947\pm0.031$ &  $1.012\pm0.079$ & $1.066\pm0.056$  \\
$z > 0.66$                 & 527 & 343 & 83 & 114 & $0.913\pm0.035$ &  $0.976\pm0.082$ & ~$1.161\pm0.072$  
\enddata
\tablecomments{Here we show the results of dividing our ensemble cluster into four redshift bins. The 
columns are: (1) the redshift intervals four each bin, (2) the total number of cluster member galaxies in 
each bin, (3) the total number of passive member galaxies in each bin, (4) the total number of post-starburst 
member galaxies in each bin, (5) the total number of star-forming member galaxies in each bin, (6) 
the velocity dispersion of passive galaxies in each bin relative to the full ensemble, (7) the 
velocity dispersion of post-starbust galaxies in each bin relative to the full ensemble, (8) 
velocity dispersion of star-forming galaxies in each bin relative to the full ensemble.}
\end{deluxetable*}

The \citet{barsanti16} analysis is essentially identical to our own except that galaxies are split into two ``red'' and 
``blue'' sub-populations based on color for the large majority of their data. These red and blue groups should translate 
approximately to isolating passive and actively star-forming galaxies, where the galaxies that we classify as post-starburst 
could be flagged as either passive (``red'') or star-forming (``blue'') galaxies, depending on the exact galaxy and color cut. 
The \citet{barsanti16} cluster sample is drawn from the literature and spans a wide range in redshift, so the red/blue 
color cut varies by cluster and the available photometry. This should have the effect of introducing some effective noise 
into the process of classifying galaxies in color space, so that there is almost certainly not a perfect mapping between 
``red'' galaxies and those which would be spectroscopically classified as passive, nor between ``blue'' and those which 
would be spectroscopically classified as star-forming. These classification differences can easily explain slight differences 
between velocity segregation measurements between our work and the \citet{barsanti16} results, though there is good 
quantitative agreement, within the uncertainties, between our two results at lower redshift, which suggests that any 
effects due to differences in classifying galaxy types are likely sub-dominant relative to the statistical uncertainties.

There is some evidence in the \citet{barsanti16} data for an easing of the velocity 
segregation between blue and red galaxies in their very broad high-z bin ($0.8 < z < 1.5$). Our measurements 
indicate that velocity segregation in our clusters between passive and star-forming galaxies is very strong in 
a bin spanning $0.66 < z < 1.08$, which could suggest that if the velocity segregation by galaxy color/spectral type 
occurs, it must be happening in clusters above $z\gtrsim1.1$. Our results are also consistent with another 
recent study by \citet{biviano16} of ten clusters in the redshift interval $0.87 \lesssim z \lesssim 1.2$ from the 
Gemini Cluster Astrophysics Spectroscopic Survey \citep[GCLASS;][]{muzzin12} in which the ratio of velocity 
dispersions of passive to star-forming galaxies is measured to be $\simeq0.88$.

\subsection{Velocity Segregation By Relative Luminosity}
\label{sec:lum_seg}

\begin{figure}
\includegraphics[scale=0.458]{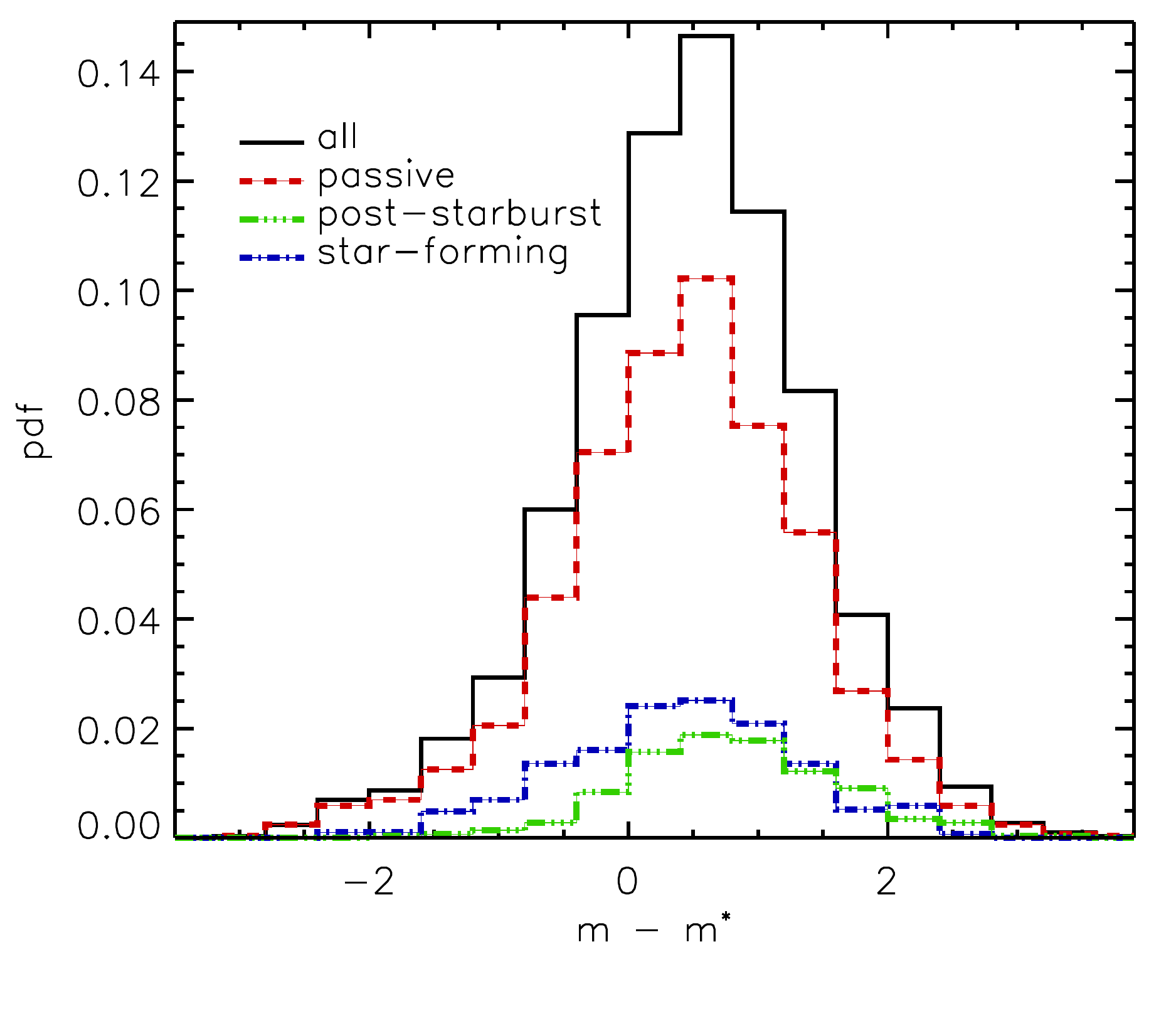}
\caption{\scriptsize{
The distribution of the brightness of the cluster galaxies analyzed here, scaled relative 
to the characteristic magnitude, $m^{*}$, at each cluster's redshift. For simplicity we use 
magnitudes in the $r-$ and $i-$bands as described in \S~\ref{sec:classification}. Here we 
show the distribution of all galaxies (solid black line), and we also highlight the contributions to the 
total sample probability distribution function from each of passive (red dash line), 
post-starburst (green dash-dot-dot-dot line), and star-forming (dash-dot-dash line) galaxies.}}
\label{fig:magnitudes}
\end{figure}

\begin{figure}
\centering
\includegraphics[scale=0.455]{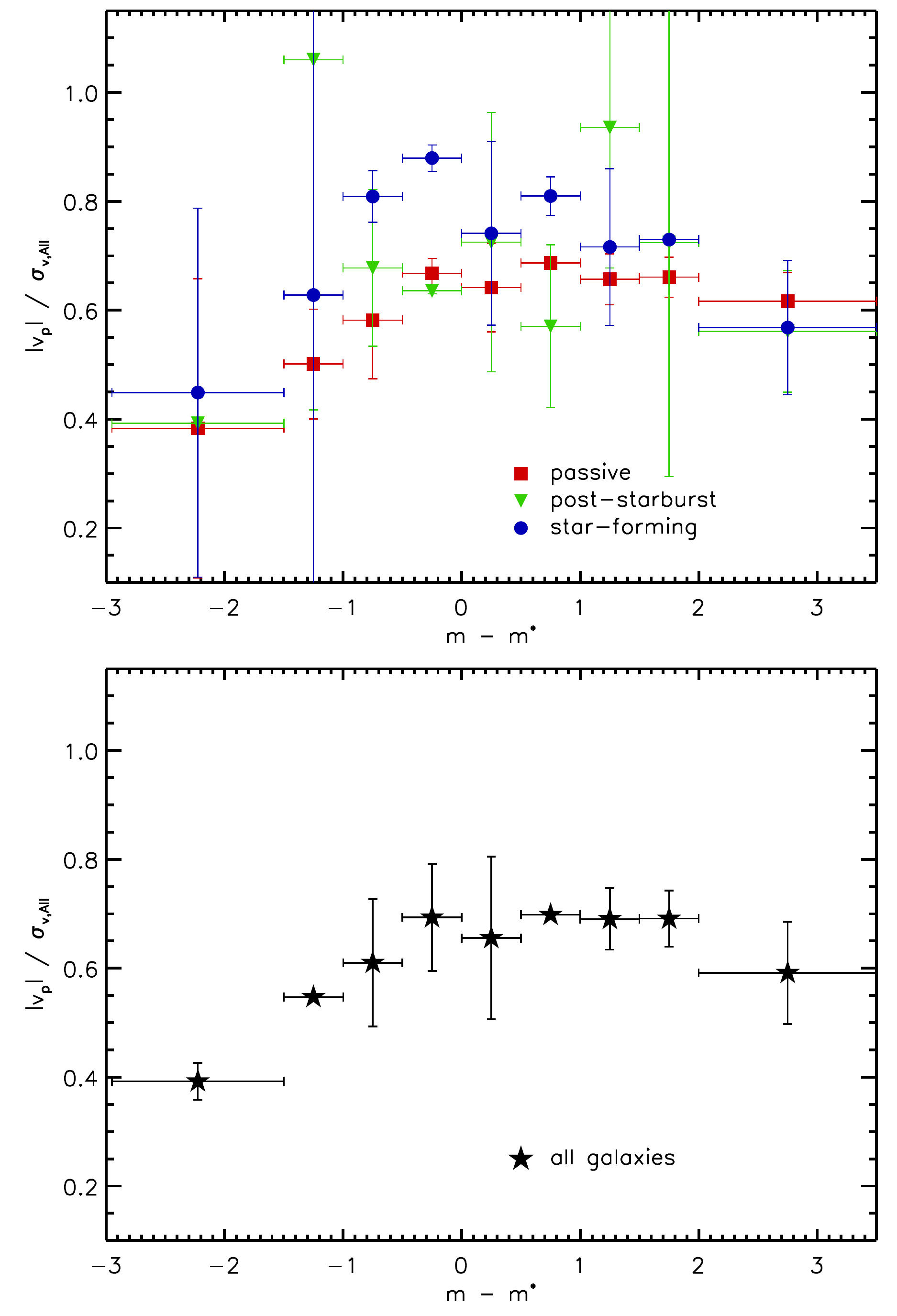}
\caption{\scriptsize{
The expectation value for the absolute peculiar velocity of cluster member galaxies as a function of 
magnitude normalized to the characteristic magnitude. In the top panel we plot the expectation 
values computed separately for the three galaxy types: passive, post-starburst, and star-forming. 
In the bottom panel we plot the expectation values computed for the full sample together, without 
consideration for spectral type. There is a distinct roll-over in which the peculiar velocities of cluster members 
drop by approximately a factor of two for galaxies brighter than $m^{*} - 0.5$, regardless of spectral type.}}
\label{fig:magbins}
\end{figure}

With photometry in hand for the majority of our galaxy spectroscopy sample we can also investigate 
velocity segregation effects as a function of galaxy luminosity. We use cluster galaxy brightness 
measurements in units relative to $m^{*}$, a standard quantity that can be easily incorporated 
into both observational data and simulated clusters \citep{cole01,cohen02,rudnick06,rudnick09}. Specifically, 
we use $m^{*}$ values computed from \citet{bruzual03} models in the same fashion as described in previous SPT 
publications \citep{high10,song12,bleem15}.  Figure~\ref{fig:magnitudes} shows the distribution of cluster member 
galaxies with brightness measurements as described in \S~\ref{sec:classification} plotted relative to the 
characteristic magnitude, $m^{*}$, for the full cluster member sample as well as the subsamples of cluster 
members of different spectral types. We use these data, in combination with the normalized peculiar velocities of 
each cluster member in the ensemble to plot the expectation value for the absolute peculiar velocity of cluster 
members as a function of brightness for all galaxies together, as well 
as for each of the passive, post-starburst, and star-forming galaxy subsets (Figure~\ref{fig:magbins}). It is clear 
that the brightest galaxies---independent of spectral type---universally prefer smaller 
absolute peculiar velocities, and that when we treat all galaxies together we see a strong drop in the absolute 
peculiar velocities of galaxies brighter than $\sim m^{*}-0.5$, while galaxies fainter than this tend to remain 
approximately flat in absolute peculiar velocity as a function of brightness. There is no statistically 
significant evidence in our data for an evolution in the presence or of velocity luminosity segregation with 
redshift. Specifically, if we split our sample into two redshift bins at $z=0.45$---which optimally balances the 
number of bright galaxies in the high and low bins---we see the strong drop in peculiar velocity for bright galaxies 
in each of the high and low redshift bins.

This same qualitative effect has been observed in low redshift 
clusters \citep{chincarini77,biviano92,mohr96,goto05,old2013,ribeiro13}.  \citet{barsanti16} also recently observed a 
similar effect in an analysis of galaxy clusters drawn from the literature. \citet{barsanti16} follow work on low-z 
clusters by \citet{biviano92} and parameterize galaxy brightnesses relative to the third brightest cluster member 
in each cluster. We find this parameterization unreliable in our own data, where we can only make probabilistic 
statements about galaxy membership beyond the galaxies for which we have spectra. Instead we choose to 
parameterize brightness relative to an observable quantity that is universally available based on standard 
measurements of the galaxy luminosity function.

Motivated by the drop in peculiar velocity that we observe in Figure~\ref{fig:magbins}, we examine the 
properties of two luminosity-based sub-populations of cluster members: those brighter than $m^{*}-0.5$ 
(``bright'') and those fainter than $m^{*}+0.5$ (``faint''). The choice of the faint galaxy sample definition is 
somewhat arbitrary, but we find that the result is insensitive to the exact cut we use here, which is unsurprising 
in light of the behavior apparent in Figure~\ref{fig:magbins}. We also apply the AD statistic in the same manner 
as before (see \S~\ref{sec:vs_by_type}) to test whether each of these sub-populations of member galaxies 
have velocity distributions that are consistent with Gaussian. Once again we point out that the formula that we 
use to convert AD statistic values into confidence levels for rejecting the null hypothesis breaks down at very 
large values of statistical significance, but it is clear that the bright and faint galaxy samples are inconsistent 
with being drawn from the same underlying velocity distribution, and that the velocity distribution of bright 
galaxies is inconsistent with a Gaussian distribution at high significance ($>6\sigma$; Table~\ref{tab:stats}). 

\begin{figure}
\centering
\includegraphics[scale=0.56]{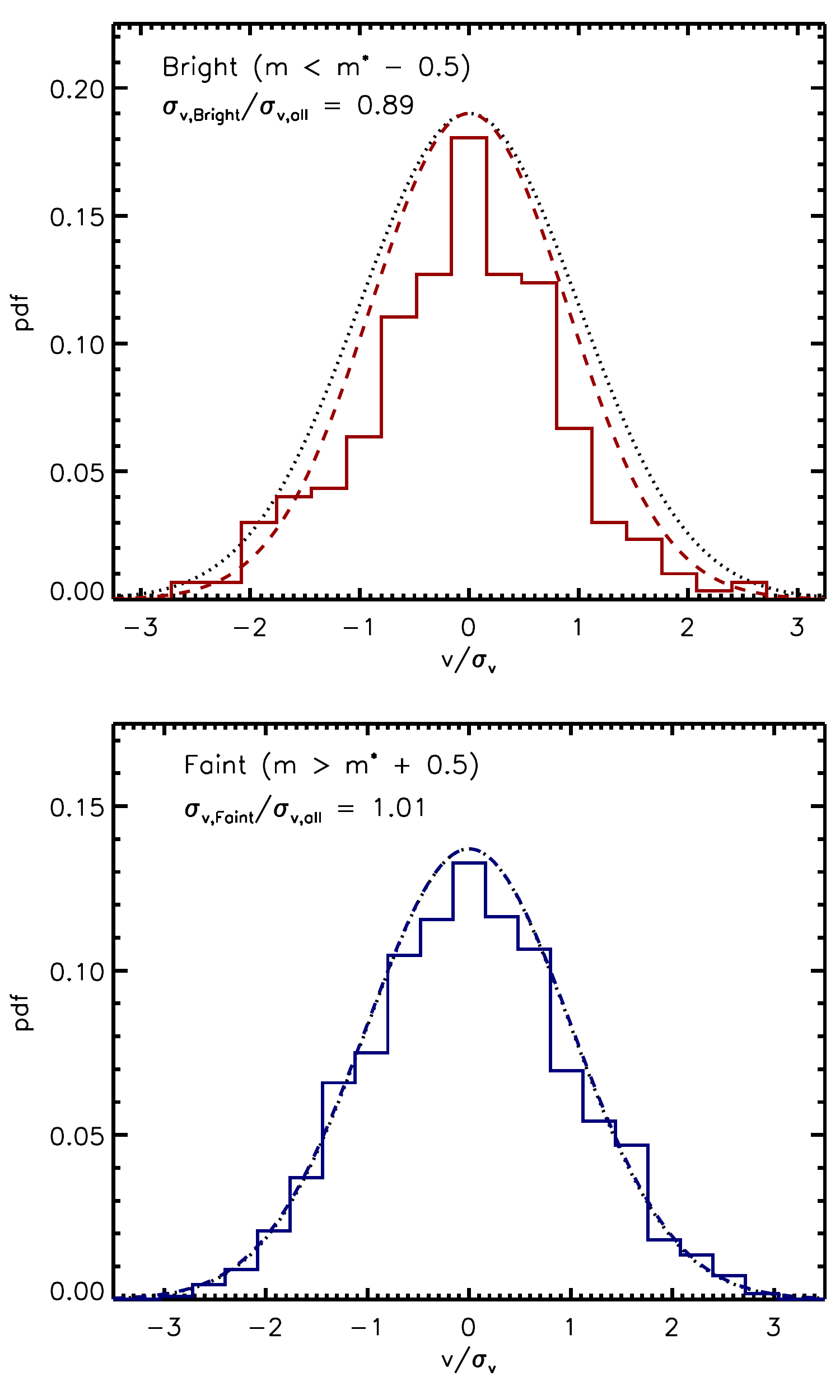}
\caption{\scriptsize{
Stacked velocity distributions for cluster member galaxies separated by brightness. Each panel includes 
a colored dashed line indicating a Gaussian distribution with the dispersion estimated from the data (solid 
histograms), as well as a black dotted line that indicates the fiducial velocity distribution, 
$\sigma_\mathrm{v,all}$, of all cluster member galaxies include in our ensemble, for comparison.
The ratio of the velocity dispersion estimated from the bright/faint subset of member galaxies to the 
dispersion estimated from the stack of all cluster member galaxies is also inset. 
\emph{Top:} The stacked velocity distribution of bright cluster members.
\emph{Bottom:} The stacked velocity distribution of faint cluster members.}}
\label{fig:mags_vdisp}
\end{figure}

Moreover, when we plot the velocity distributions of these two galaxy subsets 
(Figure~\ref{fig:mags_vdisp}) we see that the bright galaxy population has a distinctly ``peaky'' shape with an 
excess of galaxies concentrated around small peculiar velocity values and a dearth of galaxies filling out what 
would be the middle wings of a Gaussian velocity distribution. This further helps to demonstrate the divergent 
kinematics of the brightest galaxies in clusters. The behavior that we see in the peculiar velocities of bright cluster 
galaxies would appear to be direct evidence of dynamical friction effects, which should be proportional to 
the mass of a galaxy, with larger dynamical friction effects acting on brighter/more massive galaxies.

\section{Velocity Segregation and Biases in Velocity Dispersion Estimates}
\label{sec:biases}

Velocity segregation measurements are a valuable probe of cluster assembly and the link between 
environment and galaxy evolution, but they also have tremendous potential as a link between 
observations and simulations of galaxy cluster velocity dispersions as a cosmological observable. 
In this section we quantify the biases in velocity dispersion estimates by incrementally 
varying the types of galaxies that are sampled from the ensemble cluster data, and we attempt to 
compare our measured biases to those of simulated galaxy clusters. 

Specifically, we investigate the bias in the estimated velocity dispersion as a function of the 
fractions of passive/star-forming and bright/faint galaxies used. In all cases the bias that we measure 
is the bias relative to the full ensemble cluster dataset, 
and not the bias relative to the true velocity dispersion of dark matter particles in our galaxy cluster 
sample, which is, of course, unknown. We can then compare 
the biases that we measure to the same effects in simulated galaxy clusters where the true dark matter 
particle velocity dispersions are known. We advocate for the idea that comparing velocity dispersion biases 
associated with velocity segregation effects can serve as a bridge between observations and simulations 
\citep[e.g.;][]{lau10,gifford13a,saro13} of galaxy cluster velocity dispersion measurements, potentially facilitating 
a kind of cross-calibration of velocity dispersions in observations and simulations.

\begin{figure*}[t]
\centering
\includegraphics[scale=0.62]{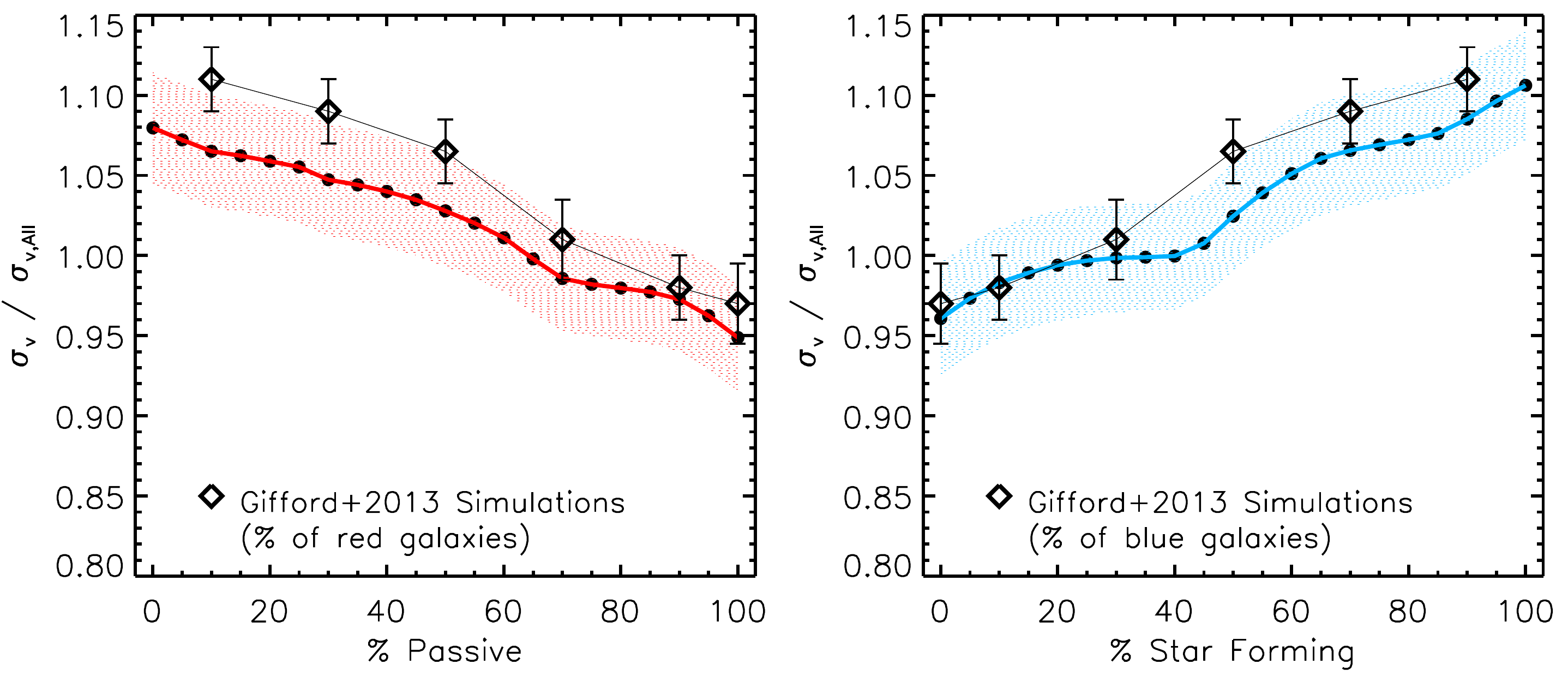}
\caption{\scriptsize{
The velocity dispersions estimated from resampling the ensemble cluster data while controlling for the fraction of passive and 
star-forming galaxies included in each resampling. In both panels we plot our results while over-plotting values (open diamonds) 
from a similar analysis of simulated galaxy clusters from \citet{gifford13a}, where the simulated galaxy clusters are populated 
with galaxies using a variety of different halo-tracking and semi-analytic prescriptions. The simulated work uses a binary red 
vs. blue galaxy classification, and so we show two panels here where the only difference between the two is in how we treat 
post-starburst galaxies in our dataset---i.e., whether they are grouped in with passive or star-forming galaxies. 
$Left:$ In this panel we vary the fraction of passive galaxies used in the resampling, using the combined 
post-starburst$+$star-forming samples to replace passive galaxies as the passive galaxy fraction is decreased.
$Right:$ In this panel we vary the fraction of star-forming galaxies used in the resampling, using the combined 
passive$+$post-starburst samples to replace star-forming galaxies as the star-forming galaxy fraction is decreased.}}
\label{fig:redvsblue}
\end{figure*}

\subsection{Velocity Dispersion vs. Fraction of Red/Blue Galaxies Used}

We first estimate velocity dispersions for resampled subsets of our ensemble galaxy cluster 
data while varying the fraction of passive and star-forming galaxies included in each resampled cluster realization. 
We essentially repeat the Monte Carlo resampling described above in \S~\ref{sec:vs_by_type}, in which each realization 
is made up of 125 galaxies drawn without replacement from the ensemble cluster, but now generate realizations in which 
we vary the fraction of the resampled cluster that is composed of a specific galaxy type. This exercise is performed with 
an eye toward comparing our results to those from simulations in the literature \citep[e.g.,][]{gifford13a}. In  
simulation-based work galaxies are often separated by color rather than precise spectral type. This means that there is 
some ambiguity about how to treat the post-starburst galaxies in our sample when comparing to other binary red vs. 
blue analyses, as we have already touched on above in \S~\ref{sec:zbins}. In light of this ambiguity we generate 
resampled datasets while varying the fraction of both the passive and star-forming fractions by regular intervals. 
When we generate resampled realizations in which we control the passive fraction we are randomly filling the 
remaining (non-passive) fraction using all of the non-passive galaxies (i.e., all post-starburst and star-forming 
galaxies). When we control the fraction of star-forming galaxies, on the other hand, then we use both post-starburst 
and passive galaxies to fill the remaining non-star-forming fraction of each realization.

The results of this resampling procedure are shown in Figure~\ref{fig:redvsblue} along with simulation results 
from \citet{gifford13a} for comparison. Qualitatively, there is good agreement between our data and simulations 
regarding the shape of the trend with red/blue fraction, independent of how post-starburst galaxies are treated. 
There is, however, a systematic offset between the normalized velocity dispersions measured in simulations and 
in our data. This offset is marginally larger when we group post-starburst galaxies in with star-forming galaxies, 
perhaps suggesting that the color cuts applied by \citet{gifford13a} preferentially sort the simulated analogs of 
post-starburst galaxies in with the passive galaxies. In the case where we treat passive and post-starburst 
galaxies together as non-star-forming, we see the closest agreement between our measurements and the 
simulations with our data lying consistently low relative to the simulations. 

The exact interpretation of the offset  is complicated, however, by the fact that the reference velocity dispersion 
($\sigma_\mathrm{v,all}$) is different in our data compared to the simulations. We use the velocity dispersion of our full 
ensemble dataset as the reference value, but clusters in simulations are plotted relative to the true underlying dark 
matter particle velocity dispersions, $\sigma_\mathrm{DM}$, a quantity that scales well with cluster virial masses in 
simulations \citep{evrard08}. We discuss this offset in more detail in \S~\ref{sec:calibrating} below.

\begin{figure*}[t]
\centering
\includegraphics[scale=0.62]{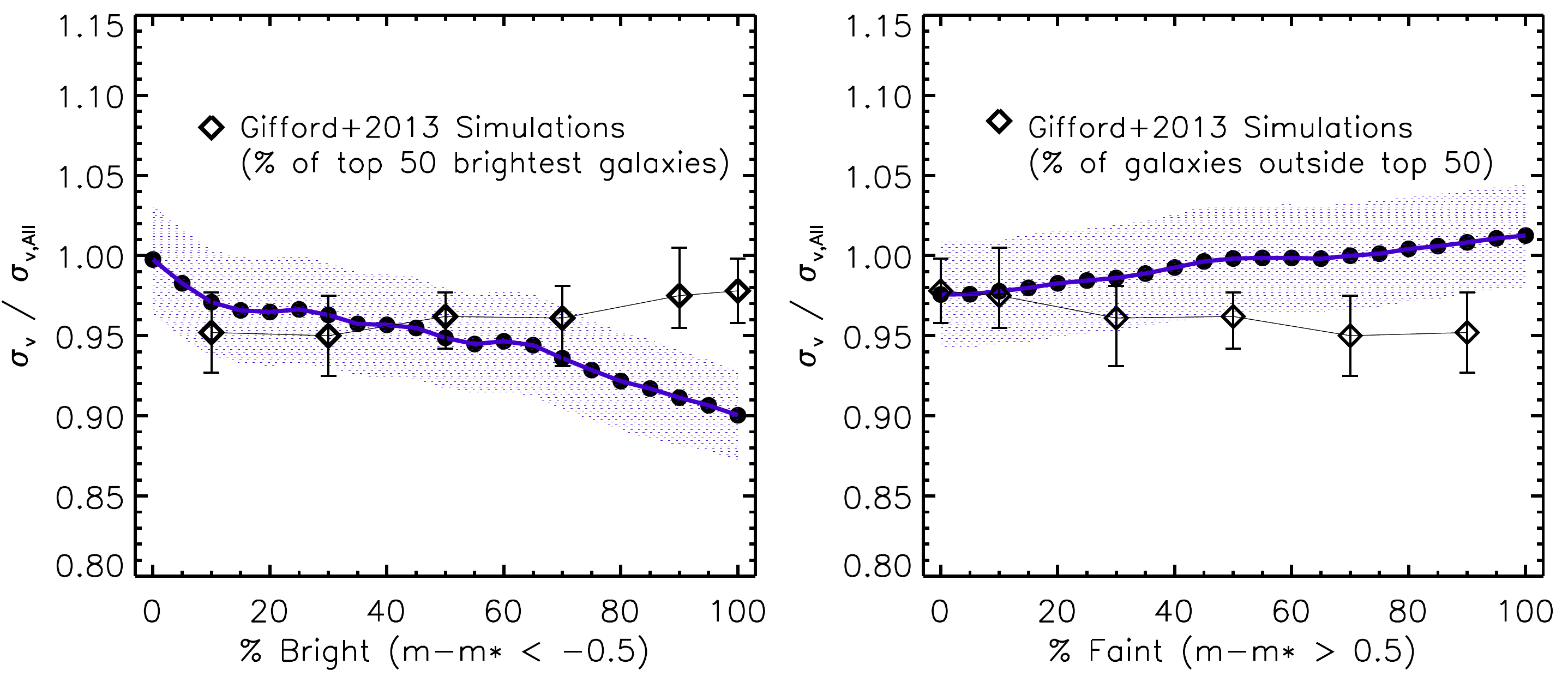}
\caption{\scriptsize{
Similar to Figure~\ref{fig:redvsblue}, but here we vary the fraction of bright and faint member galaxies used in the 
resampling, where the bright and faint samples are defined in \S~\ref{sec:lum_seg}. For comparison we over-plot 
the results (open diamonds) of a similar analysis of simulated clusters from \citet{gifford13a}, where they estimate 
velocity dispersions from resampled galaxy subsets in which they vary the fraction of the 50 brightest cluster member 
galaxies used. The simulation results plotted in the two panels are the same data, just reversed between the two panels.
$Left:$ In this panel we vary the fraction of bright galaxies used in the resampling, using all galaxies fainter than $m^{*}-0.5$ 
to replace bright galaxies as the bright galaxy fraction is decreased.
$Right:$ In this panel we vary the fraction of faint galaxies used in the resampling, using all galaxies brighter than $m^{*}+0.5$ 
to replace faint galaxies as the faint galaxy fraction is decreased.}}
\label{fig:brightvsfaint}
\end{figure*}

\subsection{Velocity Dispersion vs. Fraction of Bright/Faint Galaxies Used}

There is a significant caveat associated with making direct comparisons between our measurements and the simulation 
results from \citet{gifford13a}. Specifically, we do not have the same perfect knowledge of cluster membership that is 
available for a simulated cluster sample, and so we cannot generate a true ranking of the N brightest galaxies. Also, it is 
unlikely that we would rarely, if ever, be able to obtain spectra for each of the brightest 50 member galaxies per 
cluster---as \citet{gifford13a} do in simulations---for a real sample of clusters. It is certainly a useful exercise, however, 
to compare the relative offset in the velocity dispersion that we measure as we vary the fraction of bright vs. faint 
galaxies against predictions for a similar resampling in simulations. Our 
measurements, along with simulation results from varying the fraction of the 50 brightest members used, are shown in 
Figure~\ref{fig:brightvsfaint}. Here, again, we see the effects of the velocity segregation by galaxy luminosity, where we 
measure smaller dispersions for realizations comprised mostly or entirely of bright galaxies. Specifically, the velocity 
dispersions consistently fall as the fraction of bright galaxies increases, reaching a $\sim10$\% bias (low) for a pure 
bright galaxy dispersion (see \S~\ref{sec:lum_seg}). The simulations, however, do not exhibit the same effect, with 
dispersions biased only $\sim2\pm2$\% low when using the 50 brightest galaxies to measure the dispersion. 

Similar to the above, we also measure the velocity dispersion in resampled datasets where we vary the fraction of 
faint galaxies used, and compare this to the reverse of the simulation result. Here we see 
a weak trend with the fraction of faint galaxies used, finding velocity dispersions that are biased low by 
$\sim3\pm3$\% when using no faint galaxies, and high by $\sim1\pm3$\% low for a sample composed entirely of faint 
galaxies. We can interpret the slightly stronger bias for a sample 
using no faint galaxies as a manifestation of the velocity segregation of the brightest galaxies discussed in 
\S~\ref{sec:lum_seg}. There the ``0\% faint galaxies'' realizations include only galaxies brighter than $m^{*}+0.5$, 
which will naturally include a substantial fraction of the bright galaxy sample ($m < m^{*}-0.5$). 

\section{Discussion and Implications}
\label{sec:discussion}

There is a general consensus in the literature about the underlying astrophysical causes of 
velocity segregation effects by galaxy color or spectral/morphological type. Differences between 
passive/star-forming galaxies are understood to be related to the fact that actively star-forming galaxies 
in clusters are predominantly galaxies that are either in the process of falling into the cluster, or have 
recently done so \citep{moss77,carlberg97,goto05}. \citet{biviano04} have made a more sophisticated 
argument, claiming that star-forming cluster galaxies tend to be in approximate equilibrium with their cluster 
potentials but have orbital trajectories that are preferentially radial, while passive cluster 
galaxies have orbits that more closely resemble the random motions that would normally be associated 
with virialized test particles in a gravitational potential.

Differences in the shapes of the orbital paths of star-forming vs. passive galaxies would be a natural 
consequence of the argument that star-forming galaxies are more recently accreted, and their preferentially 
radial orbits reflect the fact that they have not yet experienced sufficient dynamical interactions within the 
cluster to acquire the more random orbits. These are all essentially differently-phrased arguments 
that star-forming member galaxies are, on average, less (or not yet) virialized, while passive member 
galaxies are virialized; these arguments are consistent with low-z studies and analyses of simulations 
aimed at understanding cluster assembling and in-fall \citep{mahajan11,haines15}.

Similarly, there is general agreement about the astrophysical effects responsible for velocity segregation 
between more and less luminous galaxies. This effect can be explained by physical 
processes such as dynamical friction (also called dynamical breaking) that cause transfers of kinetic 
energy from larger galaxies to smaller galaxies, as well as gravitational interactions that can convert 
bulk kinetic energy into internal kinetic energy via, for example, accelerating and ejecting individual 
stars during galaxy mergers \citep{sarazin86,kashlinsky87,biviano92,mahajan11}. It is mildly puzzling 
that the velocity segregation effect that we see in our data for the brightest and faintest galaxies does not 
appear in some recent simulations; we are not the first to detect this signal in bright cluster member galaxies, 
\citep{chincarini77,biviano92,mohr96,goto05,ribeiro10,old2013,barsanti16}, and the effect is seen in some 
simulations \citep[e.g.,][]{lau10,saro13}. We are wary of over-interpreting this discrepancy, because of the 
difference in our bright galaxy sample and the 50 brightest members approach used 
by \citet{gifford13a}, but it is possible that the discrepancy is related to the effects of missing baryonic/gas 
physics that are not captured in dark matter only simulations. We elect not to speculate too much about the 
pros and cons of various methods for the treatment of halos in simulated clusters, and how mock galaxies are 
placed into those halos, though it is not unreasonable to suppose that different prescriptions could generate 
significantly different results. Certainly the measurement we present here could represent useful benchmark 
test to be reproduced in future simulations. Looking forward we believe it is important that both observers 
and simulators strive to find useful quantities that can be measured in both data and simulations; it is with 
this in mind that we analyze galaxies in our sample in terms of relative luminosity, scaled by the characteristic 
magnitude, $m^{*}$.

\subsection{Velocity Segregation As A Tool To Calibrate Velocity Dispersion Estimates}
\label{sec:calibrating}

One of our primary objectives in this work is to test for the presence of systematic biases in velocity dispersion 
estimates, and to attempt to connect our results to simulations that explore the same or similar effects. In 
\S~\ref{sec:biases} we show the biases in velocity dispersions that are estimated using cluster data realizations 
in which we control the fractions of passive and star-forming galaxies. Figure~\ref{fig:redvsblue} demonstrates that 
we see the same general behavior as simulations in the change in velocity dispersion biases as a function of the 
fraction of passive and star-forming galaxies used to estimate the velocity dispersion. However, there is a small 
offset in the normalization of those biases between our data and the simulations. Independent of how we treat 
post-starburst galaxies in our analysis, whether they are grouped in with passive or star-forming 
galaxies, the velocity dispersions estimated from data as a function of galaxy type are offset low by as much as 
$\sim3$\% relative to the simulations. This offset is more pronounced for realizations comprised 
mostly or entirely of star-forming galaxies (and for those using few or no passive galaxies), reaching peak 
values of $\gtrsim3$\% when the fraction of star-forming galaxies exceeds 50\% (or when the fraction of 
passive galaxies is less than 50\%).

Here we explore the idea that these offsets may have implications for calibrating cluster velocity 
dispersion measurements relative to simulations. First, we consider the contribution of interlopers to the 
biases that we see, as it is likely that interloping star-forming galaxies in our ensemble datasets are contributing 
to some degree to the observed offset. As we have already mentioned, studies show that the inclusion of 
interloping star-forming field galaxies has the effect of driving velocity dispersion estimates to smaller values 
\citep{biviano06a}. \citet{saro13} have also 
shown that velocity dispersions estimated using passive galaxies with large interloper fractions are biased high. 
We see no evidence of our dispersions being biased high relative to simulations, suggesting that passive galaxy 
interlopers are not strongly affecting our observations.

Interloper effects are one of several different factors that can systematically bias galaxy velocity dispersions 
measurements. One interpretation that we can consider regarding the small offset that we observe in 
Figure~\ref{fig:redvsblue} is that we are effectively measuring a form of velocity bias, $b_\mathrm{v}$, 
between our galaxy velocity dispersion measurements and true underlying dark matter dispersions of our 
galaxy clusters. The standard definition of the velocity bias is 
$b_\mathrm{v} = \sigma_\mathrm{gal}/ \sigma_\mathrm{DM}$, i.e., the normalization between the true dark 
matter velocity dispersion and the velocity dispersion of cluster member galaxies.  Efforts have been 
made in the past to quantify velocity bias in several different ways, with a wide range of results 
\citep[$b_\mathrm{v} \simeq 0.95-1.3$;][]{colin00,ghigna00,diemand04,faltenbacher06,lau10,white10,saro13,wu13}, 
but to our knowledge this is the first attempt to recover an estimate of velocity bias at $z > 0.1$ by comparing velocity 
segregation effects from data and simulations. In this context our comparison suggests that velocity 
segregation measurements may provide a direct way to compare systematic biases in velocity dispersion as 
measured in real clusters and in simulations, providing a means of cross-calibrating velocity dispersion 
data and cosmological simulations.

When discussing velocity bias it is important to distinguish between velocity bias measurements in simulations, 
where both the true dark matter particle velocity dispersion and true cluster membership properties of galaxies 
are known, and in data like ours, where we do not know either. The offset that we detect in our data could 
be interpreted as a detection of what we might call the effective velocity bias, which we define as:

\begin{equation}
b_\mathrm{v,obs} = \frac{\sigma_\mathrm{gal,obs}}{\sigma_\mathrm{DM}} = \frac{\sigma_\mathrm{gal,obs}}{\sigma_\mathrm{gal,sims}} \times \frac{\sigma_\mathrm{gal,sims}}{\sigma_\mathrm{DM}} 
\end{equation}

\noindent
This is a quantity that is specific to our data and the \citet{gifford13a} simulations, in that it describes the offset that we 
observe between the velocity segregation trend between the two. A variety of different biases/effects can be folded 
into this effective velocity bias term, including dynamical friction (which should, in principle, suppress the 
peculiar velocities of all member galaxies at some level), the effects of sub-halos within the larger cluster 
halo potentials, and measurement effects such as interlopers. However, as a practical matter it is not 
necessarily important that we are able to suss out the individual factors that determine the effective velocity 
bias for our cluster spectroscopy, because $b_\mathrm{v,obs}$ is the quantity of interest if one aims to use 
velocity dispersions as accurate proxies for total cluster masses. We find that our measurements are 
offset $\sim0-3$\% low, which implies a value for $b_\mathrm{v,obs}\simeq0.97-1.0$, where the bias varies 
with different fractions of passive vs. star-forming galaxies used to measure the velocity dispersion. In 
principle, these numbers could be used to calibrate our measured dispersions to be in agreement with 
simulations---in this case specifically with the simulations analyzed by \citet{gifford13a}.

There are several important caveats to bear in mind regarding this effective velocity bias. Firstly, there is 
potentially a systematic uncertainty in our velocity segregation measurements that results from the varying 
fraction of passive vs. star-forming galaxies in our individual galaxy data (Table~\ref{tab:spectable}). This 
effect would manifest as a difference---certainly a scatter and potentially a bias---between the first ratio on the 
far right side of Equation 1 as applied to the full ensemble and the individual clusters that were used to generate 
that ensemble. The ideal dataset would avoid this systematic effect by having a constant ratio of passive to 
star-forming galaxies across all clusters, a feat that would be difficult to achieve in practice.

Secondly, different simulations measure different biases between the dispersions of dark matter and mock 
galaxies, so we might expect to recover a different offset between our data and different simulations. 
These differences would manifest as differences in the second ratio on the far right side of Equation 1. 
This represents an underlying systematic uncertainty that emerges from fundamental differences between 
simulations---e.g., volume, resolution, dark matter only vs. hydrodynamical--- as well as from how those 
simulations are populated with galaxies. We cannot address that systematic uncertainty based on our 
comparison to the \citet{gifford13a} simulations alone. That said, this is an exciting step toward quantifying this 
systematic uncertainty, which is essential to using velocity segregation to precisely calibrate dispersion 
measurements.

\section{Summary \& Conclusions}
\label{sec:conclusion}

We have analyzed spectroscopy of 4148 galaxies (2868 cluster members) in the fields of 89 massive, 
SZ-selected galaxy clusters. We detect signatures of velocity segregation as a function of both galaxy 
type and relative luminosity at high significance. We measure the velocity dispersion of star-forming galaxies 
in our cluster ensemble to be $17\pm4$\% larger than the passive galaxy population, and that this velocity 
segregation holds for our entire cluster sample, which extends to $z\sim1.1$. There is a strong drop-off in the 
average absolute peculiar velocity of cluster member galaxies brighter than $m^{*}-0.5$, with galaxies satisfying 
that criterion having a velocity dispersion $11\pm4$\% smaller than the velocity dispersion of the full 
cluster member ensemble. For the brightest galaxies we find a highly distorted velocity distribution that is 
statistically confirmed to differ from a Gaussian distribution at high significance. Finally, we compare our 
velocity segregation measurements to similar measurements in recent simulations, and see a qualitative 
agreement for passive vs. star-forming cluster member galaxies, albeit with what appears to be a systematic 
bias between the data and simulations. We consider the implications of interpreting this systematic bias 
as a detection of the effective velocity bias that describes the scaling between the observed velocity distribution 
of galaxies in clusters and the intrinsic velocity dispersion of dark matter particles in clusters, 
$b_\mathrm{v,obs}\simeq0.97-1.0$. Our result is encouraging for the prospect of using velocity segregation effects in 
galaxy cluster spectroscopy samples to calibrate velocity dispersion measurements against simulations.

\facility{Gemini-S (GMOS), Magellan:Baade (IMACS), Magellan:Clay (LDSS3), VLT:Antu (FORS2)}
    
\acknowledgments{We thank Crist{\'o}bal Sif{\'o}n for sharing reduced 1D spectra from his 2013 paper so that 
it could be included in our analysis. This work has been supported by NSF grant AST-1009012. Work at 
Argonne National Laboratory was supported under U.S. Department of Energy contract DE-AC02-06CH11357. 
DR is supported by a NASA Postdoctoral Program Senior Fellowship at the NASA Ames Research Center, 
administered by the Universities Space Research Association under contract with NASA.
Data presented here include observations from the Gemini Observatory, which is operated by the Association of 
Universities for Research in Astronomy, Inc., under a cooperative agreement with the NSF on behalf of the 
Gemini partnership: The United States, Canada, Chile, Australia, Brazil, and Argentina. Gemini data used in 
this work was taken as a part of Gemini programs awarded to PI Mohr (GS-2009B-Q-16) and PI Stubbs 
(GS-2011A-C-03, GS-2011A-C-04, GS-2011B-C-06, GS-2011B-C-33, GS-2012A-Q-04, GS-2012A-Q-37, 
GS-2012B-Q-29, GS-2012B-Q-59, GS-2013A-Q-05, GS-2013A-Q-45, GS-2013B-Q-25, GS-2013B-Q-72, 
GS-2014B-Q-31, GS-2014B-Q-64). This paper also uses spectroscopic data gathered with the 6.5-meter
Magellan Telescopes located at Las Campanas Observatory, Chile. Time was allocated through 
Harvard-CfA (PIs Bayliss, Brodwin, Foley, and Stubbs) and the Chilean National TAC (PI Clocchiatti). We 
also make use of spectroscopy from 8.1m Very Large Telescope (VLT) time granted through DDT 
(PI Carlstrom, 286.A-5021) and ESO (PI Bazin, 087.A-0843, and PI Chapman, 285.A-5034 and 088.A-0902).}

\bibliographystyle{aasjournal}
\bibliography{sptgmos}

\begin{thebibliography}{}
\expandafter\ifx\csname natexlab\endcsname\relax\def\natexlab#1{#1}\fi

\bibitem[{{Abraham} {et~al.}(1996){Abraham}, {Smecker-Hane}, {Hutchings},
  {Carlberg}, {Yee}, {Ellingson}, {Morris}, {Oke}, \& {Rigler}}]{abraham96}
{Abraham}, R.~G., {Smecker-Hane}, T.~A., {Hutchings}, J.~B., {et~al.} 1996,
  \apj, 471, 694

\bibitem[{{Allen} {et~al.}(2011){Allen}, {Evrard}, \& {Mantz}}]{allen11}
{Allen}, S.~W., {Evrard}, A.~E., \& {Mantz}, A.~B. 2011, \araa, 49, 409

\bibitem[{Bahcall {et~al.}(1991)Bahcall, Beichman, Canizares, Cronin, Heeschen,
  Houck, McKee, Noyes, Press, Sargent, Savage, Wilson, \& Wolff}]{bahcall91}
Bahcall, J.~H., Beichman, C.~A., Canizares, C., {et~al.} 1991, The Decade of
  Discovery in Astronomy and Astrophysics (Astronomy and Astrophysics Survey
  Committee, National Research Council, National Academy Press, Washington, D.
  C.)

\bibitem[{{Bahcall}(1981)}]{bahcall81}
{Bahcall}, N.~A. 1981, \apj, 247, 787

\bibitem[{Balogh {et~al.}(1999)Balogh, Babul, \& Patton}]{balogh99}
Balogh, M., Babul, A., \& Patton, D. 1999, \mnras, 307, 463

\bibitem[{{Balogh} {et~al.}(1997){Balogh}, {Morris}, {Yee}, {Carlberg}, \&
  {Ellingson}}]{balogh97}
{Balogh}, M.~L., {Morris}, S.~L., {Yee}, H.~K.~C., {Carlberg}, R.~G., \&
  {Ellingson}, E. 1997, \apjl, 488, L75

\bibitem[{{Balogh} {et~al.}(2000){Balogh}, {Navarro}, \& {Morris}}]{balogh00}
{Balogh}, M.~L., {Navarro}, J.~F., \& {Morris}, S.~L. 2000, \apj, 540, 113

\bibitem[{{Barsanti} {et~al.}(2016){Barsanti}, {Girardi}, {Biviano}, {Borgani},
  {Annunziatella}, \& {Nonino}}]{barsanti16}
{Barsanti}, S., {Girardi}, M., {Biviano}, A., {et~al.} 2016, \aap, 595, A73

\bibitem[{{Bayliss} {et~al.}(2011){Bayliss}, {Hennawi}, {Gladders}, {Koester},
  {Sharon}, {Dahle}, \& {Oguri}}]{bayliss11b}
{Bayliss}, M.~B., {Hennawi}, J.~F., {Gladders}, M.~D., {et~al.} 2011, \apjs,
  193, 8

\bibitem[{{Bayliss} {et~al.}(2014){Bayliss}, {Ashby}, {Ruel}, {Brodwin},
  {Aird}, {Bautz}, {Benson}, {Bleem}, {Bocquet}, {Carlstrom}, {Chang}, {Cho},
  {Clocchiatti}, {Crawford}, {Crites}, {Desai}, {Dobbs}, {Dudley}, {Foley},
  {Forman}, {George}, {Gettings}, {Gladders}, {Gonzalez}, {de Haan},
  {Halverson}, {High}, {Holder}, {Holzapfel}, {Hoover}, {Hrubes}, {Jones},
  {Joy}, {Keisler}, {Knox}, {Lee}, {Leitch}, {Liu}, {Lueker}, {Luong-Van},
  {Mantz}, {Marrone}, {Mawatari}, {McDonald}, {McMahon}, {Mehl}, {Meyer},
  {Miller}, {Mocanu}, {Mohr}, {Montroy}, {Murray}, {Padin}, {Plagge}, {Pryke},
  {Reichardt}, {Rest}, {Ruhl}, {Saliwanchik}, {Saro}, {Sayre}, {Schaffer},
  {Shirokoff}, {Song}, {Stalder}, {{\v S}uhada}, {Spieler}, {Stanford},
  {Staniszewski}, {Stark}, {Story}, {Stubbs}, {van Engelen}, {Vanderlinde},
  {Vieira}, {Vikhlinin}, {Williamson}, {Zahn}, \& {Zenteno}}]{bayliss14c}
{Bayliss}, M.~B., {Ashby}, M.~L.~N., {Ruel}, J., {et~al.} 2014, \apj, 794, 12

\bibitem[{{Bayliss} {et~al.}(2016){Bayliss}, {Ruel}, {Stubbs}, {Allen},
  {Applegate}, {Ashby}, {Bautz}, {Benson}, {Bleem}, {Bocquet}, {Brodwin},
  {Capasso}, {Carlstrom}, {Chang}, {Chiu}, {Cho}, {Clocchiatti}, {Crawford},
  {Crites}, {de Haan}, {Desai}, {Dietrich}, {Dobbs}, {Doucouliagos}, {Foley},
  {Forman}, {Garmire}, {George}, {Gladders}, {Gonzalez}, {Gupta}, {Halverson},
  {Hlavacek-Larrondo}, {Hoekstra}, {Holder}, {Holzapfel}, {Hou}, {Hrubes},
  {Huang}, {Jones}, {Keisler}, {Knox}, {Lee}, {Leitch}, {von der Linden},
  {Luong-Van}, {Mantz}, {Marrone}, {McDonald}, {McMahon}, {Meyer}, {Mocanu},
  {Mohr}, {Murray}, {Padin}, {Pryke}, {Rapetti}, {Reichardt}, {Rest}, {Ruhl},
  {Saliwanchik}, {Saro}, {Sayre}, {Schaffer}, {Schrabback}, {Shirokoff},
  {Song}, {Spieler}, {Stalder}, {Stanford}, {Staniszewski}, {Stark}, {Story},
  {Vanderlinde}, {Vieira}, {Vikhlinin}, {Williamson}, \& {Zenteno}}]{bayliss16}
{Bayliss}, M.~B., {Ruel}, J., {Stubbs}, C.~W., {et~al.} 2016, \apjs, 227, 3

\bibitem[{{Beers} {et~al.}(1990){Beers}, {Flynn}, \& {Gebhardt}}]{beers90}
{Beers}, T.~C., {Flynn}, K., \& {Gebhardt}, K. 1990, \aj, 100, 32

\bibitem[{{Benson} {et~al.}(2013){Benson}, {de Haan}, {Dudley}, {Reichardt},
  {Aird}, {Andersson}, {Armstrong}, {Ashby}, {Bautz}, {Bayliss}, {Bazin},
  {Bleem}, {Brodwin}, {Carlstrom}, {Chang}, {Cho}, {Clocchiatti}, {Crawford},
  {Crites}, {Desai}, {Dobbs}, {Foley}, {Forman}, {George}, {Gladders},
  {Gonzalez}, {Halverson}, {Harrington}, {High}, {Holder}, {Holzapfel},
  {Hoover}, {Hrubes}, {Jones}, {Joy}, {Keisler}, {Knox}, {Lee}, {Leitch},
  {Liu}, {Lueker}, {Luong-Van}, {Mantz}, {Marrone}, {McDonald}, {McMahon},
  {Mehl}, {Meyer}, {Mocanu}, {Mohr}, {Montroy}, {Murray}, {Natoli}, {Padin},
  {Plagge}, {Pryke}, {Rest}, {Ruel}, {Ruhl}, {Saliwanchik}, {Saro}, {Sayre},
  {Schaffer}, {Shaw}, {Shirokoff}, {Song}, {Spieler}, {Stalder},
  {Staniszewski}, {Stark}, {Story}, {Stubbs}, {Suhada}, {van Engelen},
  {Vanderlinde}, {Vieira}, {Vikhlinin}, {Williamson}, {Zahn}, \&
  {Zenteno}}]{benson13}
{Benson}, B.~A., {de Haan}, T., {Dudley}, J.~P., {et~al.} 2013, \apj, 763, 147

\bibitem[{{Biviano} {et~al.}(1992){Biviano}, {Girardi}, {Giuricin},
  {Mardirossian}, \& {Mezzetti}}]{biviano92}
{Biviano}, A., {Girardi}, M., {Giuricin}, G., {Mardirossian}, F., \&
  {Mezzetti}, M. 1992, \apj, 396, 35

\bibitem[{{Biviano} {et~al.}(1993){Biviano}, {Girardi}, {Giuricin},
  {Mardirossian}, \& {Mezzetti}}]{biviano93}
---. 1993, \apjl, 411, L13

\bibitem[{{Biviano} \& {Katgert}(2004)}]{biviano04}
{Biviano}, A., \& {Katgert}, P. 2004, \aap, 424, 779

\bibitem[{{Biviano} {et~al.}(1997){Biviano}, {Katgert}, {Mazure}, {Moles}, {den
  Hartog}, {Perea}, \& {Focardi}}]{biviano97}
{Biviano}, A., {Katgert}, P., {Mazure}, A., {et~al.} 1997, \aap, 321, 84

\bibitem[{{Biviano} {et~al.}(2002){Biviano}, {Katgert}, {Thomas}, \&
  {Adami}}]{biviano02}
{Biviano}, A., {Katgert}, P., {Thomas}, T., \& {Adami}, C. 2002, \aap, 387, 8

\bibitem[{{Biviano} {et~al.}(2006){Biviano}, {Murante}, {Borgani}, {Diaferio},
  {Dolag}, \& {Girardi}}]{biviano06a}
{Biviano}, A., {Murante}, G., {Borgani}, S., {et~al.} 2006, \aap, 456, 23

\bibitem[{{Biviano} \& {Poggianti}(2009)}]{biviano09}
{Biviano}, A., \& {Poggianti}, B.~M. 2009, \aap, 501, 419

\bibitem[{{Biviano} {et~al.}(2016){Biviano}, {van der Burg}, {Muzzin},
  {Sartoris}, {Wilson}, \& {Yee}}]{biviano16}
{Biviano}, A., {van der Burg}, R.~F.~J., {Muzzin}, A., {et~al.} 2016, \aap,
  594, A51

\bibitem[{{Biviano} {et~al.}(2013){Biviano}, {Rosati}, {Balestra}, {Mercurio},
  {Girardi}, {Nonino}, {Grillo}, {Scodeggio}, {Lemze}, {Kelson}, {Umetsu},
  {Postman}, {Zitrin}, {Czoske}, {Ettori}, {Fritz}, {Lombardi}, {Maier},
  {Medezinski}, {Mei}, {Presotto}, {Strazzullo}, {Tozzi}, {Ziegler},
  {Annunziatella}, {Bartelmann}, {Benitez}, {Bradley}, {Brescia}, {Broadhurst},
  {Coe}, {Demarco}, {Donahue}, {Ford}, {Gobat}, {Graves}, {Koekemoer},
  {Kuchner}, {Melchior}, {Meneghetti}, {Merten}, {Moustakas}, {Munari}, {Reg{\H
  o}s}, {Sartoris}, {Seitz}, \& {Zheng}}]{biviano13}
{Biviano}, A., {Rosati}, P., {Balestra}, I., {et~al.} 2013, \aap, 558, A1

\bibitem[{{Bleem} {et~al.}(2015){Bleem}, {Stalder}, {de Haan}, {Aird}, {Allen},
  {Applegate}, {Ashby}, {Bautz}, {Bayliss}, {Benson}, {Bocquet}, {Brodwin},
  {Carlstrom}, {Chang}, {Chiu}, {Cho}, {Clocchiatti}, {Crawford}, {Crites},
  {Desai}, {Dietrich}, {Dobbs}, {Foley}, {Forman}, {George}, {Gladders},
  {Gonzalez}, {Halverson}, {Hennig}, {Hoekstra}, {Holder}, {Holzapfel},
  {Hrubes}, {Jones}, {Keisler}, {Knox}, {Lee}, {Leitch}, {Liu}, {Lueker},
  {Luong-Van}, {Mantz}, {Marrone}, {McDonald}, {McMahon}, {Meyer}, {Mocanu},
  {Mohr}, {Murray}, {Padin}, {Pryke}, {Reichardt}, {Rest}, {Ruel}, {Ruhl},
  {Saliwanchik}, {Saro}, {Sayre}, {Schaffer}, {Schrabback}, {Shirokoff},
  {Song}, {Spieler}, {Stanford}, {Staniszewski}, {Stark}, {Story}, {Stubbs},
  {Vanderlinde}, {Vieira}, {Vikhlinin}, {Williamson}, {Zahn}, \&
  {Zenteno}}]{bleem15}
{Bleem}, L.~E., {Stalder}, B., {de Haan}, T., {et~al.} 2015, \apjs, 216, 27

\bibitem[{{Bocquet} {et~al.}(2015){Bocquet}, {Saro}, {Mohr}, {Aird}, {Ashby},
  {Bautz}, {Bayliss}, {Bazin}, {Benson}, {Bleem}, {Brodwin}, {Carlstrom},
  {Chang}, {Chiu}, {Cho}, {Clocchiatti}, {Crawford}, {Crites}, {Desai}, {de
  Haan}, {Dietrich}, {Dobbs}, {Foley}, {Forman}, {Gangkofner}, {George},
  {Gladders}, {Gonzalez}, {Halverson}, {Hennig}, {Hlavacek-Larrondo}, {Holder},
  {Holzapfel}, {Hrubes}, {Jones}, {Keisler}, {Knox}, {Lee}, {Leitch}, {Liu},
  {Lueker}, {Luong-Van}, {Marrone}, {McDonald}, {McMahon}, {Meyer}, {Mocanu},
  {Murray}, {Padin}, {Pryke}, {Reichardt}, {Rest}, {Ruel}, {Ruhl},
  {Saliwanchik}, {Sayre}, {Schaffer}, {Shirokoff}, {Spieler}, {Stalder},
  {Stanford}, {Staniszewski}, {Stark}, {Story}, {Stubbs}, {Vanderlinde},
  {Vieira}, {Vikhlinin}, {Williamson}, {Zahn}, \& {Zenteno}}]{bocquet15}
{Bocquet}, S., {Saro}, A., {Mohr}, J.~J., {et~al.} 2015, \apj, 799, 214

\bibitem[{{Borgani} {et~al.}(1997){Borgani}, {da Costa}, {Freudling},
  {Giovanelli}, {Haynes}, {Salzer}, \& {Wegner}}]{borgani97}
{Borgani}, S., {da Costa}, L.~N., {Freudling}, W., {et~al.} 1997, \apjl, 482,
  L121

\bibitem[{{Bothun}(1982)}]{bothun82}
{Bothun}, G.~D. 1982, \apjs, 50, 39

\bibitem[{{Brodwin} {et~al.}(2010){Brodwin}, {Ruel}, {Ade}, {Aird},
  {Andersson}, {Ashby}, {Bautz}, {Bazin}, {Benson}, {Bleem}, {Carlstrom},
  {Chang}, {Crawford}, {Crites}, {de Haan}, {Desai}, {Dobbs}, {Dudley},
  {Fazio}, {Foley}, {Forman}, {Garmire}, {George}, {Gladders}, {Gonzalez},
  {Halverson}, {High}, {Holder}, {Holzapfel}, {Hrubes}, {Jones}, {Joy},
  {Keisler}, {Knox}, {Lee}, {Leitch}, {Lueker}, {Marrone}, {McMahon}, {Mehl},
  {Meyer}, {Mohr}, {Montroy}, {Murray}, {Padin}, {Plagge}, {Pryke},
  {Reichardt}, {Rest}, {Ruhl}, {Schaffer}, {Shaw}, {Shirokoff}, {Song},
  {Spieler}, {Stalder}, {Stanford}, {Staniszewski}, {Stark}, {Stubbs},
  {Vanderlinde}, {Vieira}, {Vikhlinin}, {Williamson}, {Yang}, {Zahn}, \&
  {Zenteno}}]{brodwin10}
{Brodwin}, M., {Ruel}, J., {Ade}, P.~A.~R., {et~al.} 2010, \apj, 721, 90

\bibitem[{{Brodwin} {et~al.}(2013){Brodwin}, {Stanford}, {Gonzalez}, {Zeimann},
  {Snyder}, {Mancone}, {Pope}, {Eisenhardt}, {Stern}, {Alberts}, {Ashby},
  {Brown}, {Chary}, {Dey}, {Galametz}, {Gettings}, {Jannuzi}, {Miller},
  {Moustakas}, \& {Moustakas}}]{brodwin13}
{Brodwin}, M., {Stanford}, S.~A., {Gonzalez}, A.~H., {et~al.} 2013, \apj, 779,
  138

\bibitem[{{Bruzual} \& {Charlot}(2003)}]{bruzual03}
{Bruzual}, G., \& {Charlot}, S. 2003, \mnras, 344, 1000

\bibitem[{{Carlberg} {et~al.}(1997{\natexlab{a}}){Carlberg}, {Yee},
  {Ellingson}, {Morris}, {Abraham}, {Gravel}, {Pritchet}, {Smecker-Hane},
  {Hartwick}, {Hesser}, {Hutchings}, \& {Oke}}]{carlberg97}
{Carlberg}, R.~G., {Yee}, H.~K.~C., {Ellingson}, E., {et~al.}
  1997{\natexlab{a}}, \apjl, 485, L13+

\bibitem[{{Carlberg} {et~al.}(1997{\natexlab{b}}){Carlberg}, {Yee},
  {Ellingson}, {Morris}, {Abraham}, {Gravel}, {Pritchet}, {Smecker-Hane},
  {Hartwick}, {Hesser}, {Hutchings}, \& {Oke}}]{carlberg97b}
---. 1997{\natexlab{b}}, \apjl, 476, L7

\bibitem[{{Carlstrom} {et~al.}(2011){Carlstrom}, {Ade}, {Aird}, {Benson},
  {Bleem}, {Busetti}, {Chang}, {Chauvin}, {Cho}, {Crawford}, {Crites}, {Dobbs},
  {Halverson}, {Heimsath}, {Holzapfel}, {Hrubes}, {Joy}, {Keisler}, {Lanting},
  {Lee}, {Leitch}, {Leong}, {Lu}, {Lueker}, {Luong-van}, {McMahon}, {Mehl},
  {Meyer}, {Mohr}, {Montroy}, {Padin}, {Plagge}, {Pryke}, {Ruhl}, {Schaffer},
  {Schwan}, {Shirokoff}, {Spieler}, {Staniszewski}, {Stark}, {Tucker},
  {Vanderlinde}, {Vieira}, \& {Williamson}}]{carlstrom11}
{Carlstrom}, J.~E., {Ade}, P.~A.~R., {Aird}, K.~A., {et~al.} 2011, \pasp, 123,
  568

\bibitem[{{Chincarini} \& {Rood}(1977)}]{chincarini77}
{Chincarini}, G., \& {Rood}, H.~J. 1977, \apj, 214, 351

\bibitem[{{Cohen}(2002)}]{cohen02}
{Cohen}, J.~G. 2002, \apj, 567, 672

\bibitem[{{Cole} {et~al.}(2001){Cole}, {Norberg}, {Baugh}, {Frenk},
  {Bland-Hawthorn}, {Bridges}, {Cannon}, {Colless}, {Collins}, {Couch},
  {Cross}, {Dalton}, {De Propris}, {Driver}, {Efstathiou}, {Ellis},
  {Glazebrook}, {Jackson}, {Lahav}, {Lewis}, {Lumsden}, {Maddox}, {Madgwick},
  {Peacock}, {Peterson}, {Sutherland}, \& {Taylor}}]{cole01}
{Cole}, S., {Norberg}, P., {Baugh}, C.~M., {et~al.} 2001, \mnras, 326, 255

\bibitem[{{Col{\'{\i}}n} {et~al.}(2000){Col{\'{\i}}n}, {Klypin}, \&
  {Kravtsov}}]{colin00}
{Col{\'{\i}}n}, P., {Klypin}, A.~A., \& {Kravtsov}, A.~V. 2000, \apj, 539, 561

\bibitem[{{Colless} \& {Dunn}(1996)}]{colless96}
{Colless}, M., \& {Dunn}, A.~M. 1996, \apj, 458, 435

\bibitem[{{Crawford} {et~al.}(2014){Crawford}, {Wirth}, \&
  {Bershady}}]{crawford14}
{Crawford}, S.~M., {Wirth}, G.~D., \& {Bershady}, M.~A. 2014, \apj, 786, 30

\bibitem[{{de Haan} {et~al.}(2016){de Haan}, {Benson}, {Bleem}, {Allen},
  {Applegate}, {Ashby}, {Bautz}, {Bayliss}, {Bocquet}, {Brodwin}, {Carlstrom},
  {Chang}, {Chiu}, {Cho}, {Clocchiatti}, {Crawford}, {Crites}, {Desai},
  {Dietrich}, {Dobbs}, {Doucouliagos}, {Foley}, {Forman}, {Garmire}, {George},
  {Gladders}, {Gonzalez}, {Gupta}, {Halverson}, {Hlavacek-Larrondo},
  {Hoekstra}, {Holder}, {Holzapfel}, {Hou}, {Hrubes}, {Huang}, {Jones},
  {Keisler}, {Knox}, {Lee}, {Leitch}, {von der Linden}, {Luong-Van}, {Mantz},
  {Marrone}, {McDonald}, {McMahon}, {Meyer}, {Mocanu}, {Mohr}, {Murray},
  {Padin}, {Pryke}, {Rapetti}, {Reichardt}, {Rest}, {Ruel}, {Ruhl},
  {Saliwanchik}, {Saro}, {Sayre}, {Schaffer}, {Schrabback}, {Shirokoff},
  {Song}, {Spieler}, {Stalder}, {Stanford}, {Staniszewski}, {Stark}, {Story},
  {Stubbs}, {Vanderlinde}, {Vieira}, {Vikhlinin}, {Williamson}, \&
  {Zenteno}}]{dehaan16}
{de Haan}, T., {Benson}, B.~A., {Bleem}, L.~E., {et~al.} 2016, \apj, 832, 95

\bibitem[{{Diemand} {et~al.}(2004){Diemand}, {Moore}, \& {Stadel}}]{diemand04}
{Diemand}, J., {Moore}, B., \& {Stadel}, J. 2004, \mnras, 352, 535

\bibitem[{{Diemer} {et~al.}(2013){Diemer}, {Kravtsov}, \& {More}}]{diemer13}
{Diemer}, B., {Kravtsov}, A.~V., \& {More}, S. 2013, \apj, 779, 159

\bibitem[{{Dom{\'{\i}}nguez} {et~al.}(2001){Dom{\'{\i}}nguez}, {Muriel}, \&
  {Lambas}}]{dominguez01}
{Dom{\'{\i}}nguez}, M., {Muriel}, H., \& {Lambas}, D.~G. 2001, \aj, 121, 1266

\bibitem[{{Dressler}(1980)}]{dressler80}
{Dressler}, A. 1980, \apj, 236, 351

\bibitem[{{Dressler} {et~al.}(2006){Dressler}, {Hare}, {Bigelow}, \&
  {Osip}}]{dressler06}
{Dressler}, A., {Hare}, T., {Bigelow}, B.~C., \& {Osip}, D.~J. 2006, in Society
  of Photo-Optical Instrumentation Engineers (SPIE) Conference Series, Vol.
  6269, Society of Photo-Optical Instrumentation Engineers (SPIE) Conference
  Series

\bibitem[{{Dressler} {et~al.}(1999){Dressler}, {Smail}, {Poggianti}, {Butcher},
  {Couch}, {Ellis}, \& {Oemler}}]{dressler99}
{Dressler}, A., {Smail}, I., {Poggianti}, B.~M., {et~al.} 1999, \apjs, 122, 51

\bibitem[{{Evrard} {et~al.}(2008){Evrard}, {Bialek}, {Busha}, {White}, {Habib},
  {Heitmann}, {Warren}, {Rasia}, {Tormen}, {Moscardini}, {Power}, {Jenkins},
  {Gao}, {Frenk}, {Springel}, {White}, \& {Diemand}}]{evrard08}
{Evrard}, A.~E., {Bialek}, J., {Busha}, M., {et~al.} 2008, \apj, 672, 122

\bibitem[{{Fadda} {et~al.}(1996){Fadda}, {Girardi}, {Giuricin}, {Mardirossian},
  \& {Mezzetti}}]{fadda96}
{Fadda}, D., {Girardi}, M., {Giuricin}, G., {Mardirossian}, F., \& {Mezzetti},
  M. 1996, \apj, 473, 670

\bibitem[{{Faltenbacher} \& {Diemand}(2006)}]{faltenbacher06}
{Faltenbacher}, A., \& {Diemand}, J. 2006, \mnras, 369, 1698

\bibitem[{{Geller} {et~al.}(2013){Geller}, {Diaferio}, {Rines}, \&
  {Serra}}]{geller13}
{Geller}, M.~J., {Diaferio}, A., {Rines}, K.~J., \& {Serra}, A.~L. 2013, \apj,
  764, 58

\bibitem[{{Gerken} {et~al.}(2004){Gerken}, {Ziegler}, {Balogh}, {Gilbank},
  {Fritz}, \& {J{\"a}ger}}]{gerken04}
{Gerken}, B., {Ziegler}, B., {Balogh}, M., {et~al.} 2004, \aap, 421, 59

\bibitem[{{Ghigna} {et~al.}(2000){Ghigna}, {Moore}, {Governato}, {Lake},
  {Quinn}, \& {Stadel}}]{ghigna00}
{Ghigna}, S., {Moore}, B., {Governato}, F., {et~al.} 2000, \apj, 544, 616

\bibitem[{{Gifford} {et~al.}(2013){Gifford}, {Miller}, \& {Kern}}]{gifford13a}
{Gifford}, D., {Miller}, C., \& {Kern}, N. 2013, \apj, 773, 116

\bibitem[{{Girardi} {et~al.}(1993){Girardi}, {Biviano}, {Giuricin},
  {Mardirossian}, \& {Mezzetti}}]{girardi93}
{Girardi}, M., {Biviano}, A., {Giuricin}, G., {Mardirossian}, F., \&
  {Mezzetti}, M. 1993, \apj, 404, 38

\bibitem[{{Girardi} {et~al.}(1996){Girardi}, {Fadda}, {Giuricin},
  {Mardirossian}, {Mezzetti}, \& {Biviano}}]{girardi96}
{Girardi}, M., {Fadda}, D., {Giuricin}, G., {et~al.} 1996, \apj, 457, 61

\bibitem[{{Girardi} {et~al.}(2015){Girardi}, {Mercurio}, {Balestra}, {Nonino},
  {Biviano}, {Grillo}, {Rosati}, {Annunziatella}, {Demarco}, {Fritz}, {Gobat},
  {Lemze}, {Presotto}, {Scodeggio}, {Tozzi}, {Bartosch Caminha}, {Brescia},
  {Coe}, {Kelson}, {Koekemoer}, {Lombardi}, {Medezinski}, {Postman},
  {Sartoris}, {Umetsu}, {Zitrin}, {Boschin}, {Czoske}, {De Lucia}, {Kuchner},
  {Maier}, {Meneghetti}, {Monaco}, {Monna}, {Munari}, {Seitz}, {Verdugo}, \&
  {Ziegler}}]{girardi15}
{Girardi}, M., {Mercurio}, A., {Balestra}, I., {et~al.} 2015, \aap, 579, A4

\bibitem[{{Goto}(2005)}]{goto05}
{Goto}, T. 2005, \mnras, 359, 1415

\bibitem[{{Haines} {et~al.}(2015){Haines}, {Pereira}, {Smith}, {Egami},
  {Babul}, {Finoguenov}, {Ziparo}, {McGee}, {Rawle}, {Okabe}, \&
  {Moran}}]{haines15}
{Haines}, C.~P., {Pereira}, M.~J., {Smith}, G.~P., {et~al.} 2015, \apj, 806,
  101

\bibitem[{{Hamuy} {et~al.}(1992){Hamuy}, {Walker}, {Suntzeff}, {Gigoux},
  {Heathcote}, \& {Phillips}}]{hamuy1992}
{Hamuy}, M., {Walker}, A.~R., {Suntzeff}, N.~B., {et~al.} 1992, \pasp, 104, 533

\bibitem[{{Hasselfield} {et~al.}(2013){Hasselfield}, {Hilton}, {Marriage},
  {Addison}, {Barrientos}, {Battaglia}, {Battistelli}, {Bond}, {Crichton},
  {Das}, {Devlin}, {Dicker}, {Dunkley}, {D{\"u}nner}, {Fowler}, {Gralla},
  {Hajian}, {Halpern}, {Hincks}, {Hlozek}, {Hughes}, {Infante}, {Irwin},
  {Kosowsky}, {Marsden}, {Menanteau}, {Moodley}, {Niemack}, {Nolta}, {Page},
  {Partridge}, {Reese}, {Schmitt}, {Sehgal}, {Sherwin}, {Sievers}, {Sif{\'o}n},
  {Spergel}, {Staggs}, {Swetz}, {Switzer}, {Thornton}, {Trac}, \&
  {Wollack}}]{hasselfield2013}
{Hasselfield}, M., {Hilton}, M., {Marriage}, T.~A., {et~al.} 2013, \jcap, 7, 8

\bibitem[{{Hern{\'a}ndez-Fern{\'a}ndez}
  {et~al.}(2014){Hern{\'a}ndez-Fern{\'a}ndez}, {Haines}, {Diaferio},
  {Iglesias-P{\'a}ramo}, {Mendes de Oliveira}, \& {Vilchez}}]{hernandez14}
{Hern{\'a}ndez-Fern{\'a}ndez}, J.~D., {Haines}, C.~P., {Diaferio}, A., {et~al.}
  2014, \mnras, 438, 2186

\bibitem[{{High} {et~al.}(2010){High}, {Stalder}, {Song}, {Ade}, {Aird},
  {Allam}, {Armstrong}, {Barkhouse}, {Benson}, {Bertin}, {Bhattacharya},
  {Bleem}, {Brodwin}, {Buckley-Geer}, {Carlstrom}, {Challis}, {Chang},
  {Crawford}, {Crites}, {de Haan}, {Desai}, {Dobbs}, {Dudley}, {Foley},
  {George}, {Gladders}, {Halverson}, {Hamuy}, {Hansen}, {Holder}, {Holzapfel},
  {Hrubes}, {Joy}, {Keisler}, {Lee}, {Leitch}, {Lin}, {Lin}, {Loehr}, {Lueker},
  {Marrone}, {McMahon}, {Mehl}, {Meyer}, {Mohr}, {Montroy}, {Morell}, {Ngeow},
  {Padin}, {Plagge}, {Pryke}, {Reichardt}, {Rest}, {Ruel}, {Ruhl}, {Schaffer},
  {Shaw}, {Shirokoff}, {Smith}, {Spieler}, {Staniszewski}, {Stark}, {Stubbs},
  {Tucker}, {Vanderlinde}, {Vieira}, {Williamson}, {Wood-Vasey}, {Yang},
  {Zahn}, \& {Zenteno}}]{high10}
{High}, F.~W., {Stalder}, B., {Song}, J., {et~al.} 2010, \apj, 723, 1736

\bibitem[{{Hou} {et~al.}(2009){Hou}, {Parker}, {Harris}, \& {Wilman}}]{hou2009}
{Hou}, A., {Parker}, L.~C., {Harris}, W.~E., \& {Wilman}, D.~J. 2009, \apj,
  702, 1199

\bibitem[{{Hubble}(1926)}]{hubble26}
{Hubble}, E.~P. 1926, \apj, 64, doi:10.1086/143018

\bibitem[{{Hwang} \& {Lee}(2008)}]{hwang08}
{Hwang}, H.~S., \& {Lee}, M.~G. 2008, \apj, 676, 218

\bibitem[{{Jordi} {et~al.}(2006){Jordi}, {Grebel}, \& {Ammon}}]{jordi2006}
{Jordi}, K., {Grebel}, E.~K., \& {Ammon}, K. 2006, \aap, 460, 339

\bibitem[{{Kashlinsky}(1987)}]{kashlinsky87}
{Kashlinsky}, A. 1987, \apj, 312, 497

\bibitem[{{Kasun} \& {Evrard}(2005)}]{kasun05}
{Kasun}, S.~F., \& {Evrard}, A.~E. 2005, \apj, 629, 781

\bibitem[{{Katgert} {et~al.}(1996){Katgert}, {Mazure}, {Perea}, {den Hartog},
  {Moles}, {Le Fevre}, {Dubath}, {Focardi}, {Rhee}, {Jones}, {Escalera},
  {Biviano}, {Gerbal}, \& {Giuricin}}]{katgert96}
{Katgert}, P., {Mazure}, A., {Perea}, J., {et~al.} 1996, \aap, 310, 8

\bibitem[{{Kelson}(2003)}]{kelson03}
{Kelson}, D.~D. 2003, \pasp, 115, 688

\bibitem[{{Kent} \& {Gunn}(1982)}]{kent82}
{Kent}, S.~M., \& {Gunn}, J.~E. 1982, \aj, 87, 945

\bibitem[{{Kurtz} \& {Mink}(1998)}]{kurtz98}
{Kurtz}, M.~J., \& {Mink}, D.~J. 1998, \pasp, 110, 934

\bibitem[{{Larson} {et~al.}(1980){Larson}, {Tinsley}, \& {Caldwell}}]{larson80}
{Larson}, R.~B., {Tinsley}, B.~M., \& {Caldwell}, C.~N. 1980, \apj, 237, 692

\bibitem[{{Lau} {et~al.}(2010){Lau}, {Nagai}, \& {Kravtsov}}]{lau10}
{Lau}, E.~T., {Nagai}, D., \& {Kravtsov}, A.~V. 2010, \apj, 708, 1419

\bibitem[{{Mahajan} {et~al.}(2011){Mahajan}, {Mamon}, \&
  {Raychaudhury}}]{mahajan11}
{Mahajan}, S., {Mamon}, G.~A., \& {Raychaudhury}, S. 2011, \mnras, 416, 2882

\bibitem[{{Majumdar} \& {Mohr}(2003)}]{majumdar03}
{Majumdar}, S., \& {Mohr}, J.~J. 2003, \apj, 585, 603

\bibitem[{{Majumdar} \& {Mohr}(2004)}]{majumdar04}
---. 2004, \apj, 613, 41

\bibitem[{{Marriage} {et~al.}(2011){Marriage}, {Acquaviva}, {Ade}, {Aguirre},
  {Amiri}, {Appel}, {Barrientos}, {Battistelli}, {Bond}, {Brown}, {Burger},
  {Chervenak}, {Das}, {Devlin}, {Dicker}, {Bertrand Doriese}, {Dunkley},
  {D{\"u}nner}, {Essinger-Hileman}, {Fisher}, {Fowler}, {Hajian}, {Halpern},
  {Hasselfield}, {Hern{\'a}ndez-Monteagudo}, {Hilton}, {Hilton}, {Hincks},
  {Hlozek}, {Huffenberger}, {Handel Hughes}, {Hughes}, {Infante}, {Irwin},
  {Baptiste Juin}, {Kaul}, {Klein}, {Kosowsky}, {Lau}, {Limon}, {Lin},
  {Lupton}, {Marsden}, {Martocci}, {Mauskopf}, {Menanteau}, {Moodley},
  {Moseley}, {Netterfield}, {Niemack}, {Nolta}, {Page}, {Parker}, {Partridge},
  {Quintana}, {Reese}, {Reid}, {Sehgal}, {Sherwin}, {Sievers}, {Spergel},
  {Staggs}, {Swetz}, {Switzer}, {Thornton}, {Trac}, {Tucker}, {Warne},
  {Wilson}, {Wollack}, \& {Zhao}}]{marriage11b}
{Marriage}, T.~A., {Acquaviva}, V., {Ade}, P.~A.~R., {et~al.} 2011, \apj, 737,
  61

\bibitem[{{Melnick} \& {Sargent}(1977)}]{melnick77}
{Melnick}, J., \& {Sargent}, W.~L.~W. 1977, \apj, 215, 401

\bibitem[{{Mohr} {et~al.}(1996){Mohr}, {Geller}, \& {Wegner}}]{mohr96}
{Mohr}, J.~J., {Geller}, M.~J., \& {Wegner}, G. 1996, \aj, 112, 1816

\bibitem[{{Moss} \& {Dickens}(1977)}]{moss77}
{Moss}, C., \& {Dickens}, R.~J. 1977, \mnras, 178, 701

\bibitem[{{Munari} {et~al.}(2013){Munari}, {Biviano}, {Borgani}, {Murante}, \&
  {Fabjan}}]{munari13}
{Munari}, E., {Biviano}, A., {Borgani}, S., {Murante}, G., \& {Fabjan}, D.
  2013, \mnras, 430, 2638

\bibitem[{{Muzzin} {et~al.}(2012){Muzzin}, {Wilson}, {Yee}, {Gilbank},
  {Hoekstra}, {Demarco}, {Balogh}, {van Dokkum}, {Franx}, {Ellingson}, {Hicks},
  {Nantais}, {Noble}, {Lacy}, {Lidman}, {Rettura}, {Surace}, \&
  {Webb}}]{muzzin12}
{Muzzin}, A., {Wilson}, G., {Yee}, H.~K.~C., {et~al.} 2012, \apj, 746, 188

\bibitem[{{Muzzin} {et~al.}(2014){Muzzin}, {van der Burg}, {McGee}, {Balogh},
  {Franx}, {Hoekstra}, {Hudson}, {Noble}, {Taranu}, {Webb}, {Wilson}, \&
  {Yee}}]{muzzin14}
{Muzzin}, A., {van der Burg}, R.~F.~J., {McGee}, S.~L., {et~al.} 2014, \apj,
  796, 65

\bibitem[{{Noh} \& {Cohn}(2012)}]{noh12}
{Noh}, Y., \& {Cohn}, J.~D. 2012, \mnras, 426, 1829

\bibitem[{{Old} {et~al.}(2013){Old}, {Gray}, \& {Pearce}}]{old2013}
{Old}, L., {Gray}, M.~E., \& {Pearce}, F.~R. 2013, \mnras, 434, 2606

\bibitem[{{Owers} {et~al.}(2011){Owers}, {Randall}, {Nulsen}, {Couch}, {David},
  \& {Kempner}}]{owers11}
{Owers}, M.~S., {Randall}, S.~W., {Nulsen}, P.~E.~J., {et~al.} 2011, \apj, 728,
  27

\bibitem[{{Planck Collaboration} {et~al.}(2013){Planck Collaboration}, {Ade},
  {Aghanim}, {Armitage-Caplan}, {Arnaud}, {Ashdown}, {Atrio-Barandela},
  {Aumont}, {Baccigalupi}, {Banday}, \& et~al.}]{planck13-XX}
{Planck Collaboration}, {Ade}, P.~A.~R., {Aghanim}, N., {et~al.} 2013, ArXiv
  e-prints, arXiv:1303.5080

\bibitem[{{Quintana} {et~al.}(2000){Quintana}, {Carrasco}, \&
  {Reisenegger}}]{quintana00}
{Quintana}, H., {Carrasco}, E.~R., \& {Reisenegger}, A. 2000, \aj, 120, 511

\bibitem[{{Reichardt} {et~al.}(2013){Reichardt}, {Stalder}, {Bleem}, {Montroy},
  {Aird}, {Andersson}, {Armstrong}, {Ashby}, {Bautz}, {Bayliss}, {Bazin},
  {Benson}, {Brodwin}, {Carlstrom}, {Chang}, {Cho}, {Clocchiatti}, {Crawford},
  {Crites}, {de Haan}, {Desai}, {Dobbs}, {Dudley}, {Foley}, {Forman}, {George},
  {Gladders}, {Gonzalez}, {Halverson}, {Harrington}, {High}, {Holder},
  {Holzapfel}, {Hoover}, {Hrubes}, {Jones}, {Joy}, {Keisler}, {Knox}, {Lee},
  {Leitch}, {Liu}, {Lueker}, {Luong-Van}, {Mantz}, {Marrone}, {McDonald},
  {McMahon}, {Mehl}, {Meyer}, {Mocanu}, {Mohr}, {Murray}, {Natoli}, {Padin},
  {Plagge}, {Pryke}, {Rest}, {Ruel}, {Ruhl}, {Saliwanchik}, {Saro}, {Sayre},
  {Schaffer}, {Shaw}, {Shirokoff}, {Song}, {Spieler}, {Staniszewski}, {Stark},
  {Story}, {Stubbs}, {{\v S}uhada}, {van Engelen}, {Vanderlinde}, {Vieira},
  {Vikhlinin}, {Williamson}, {Zahn}, \& {Zenteno}}]{reichardt13}
{Reichardt}, C.~L., {Stalder}, B., {Bleem}, L.~E., {et~al.} 2013, \apj, 763,
  127

\bibitem[{{Ribeiro} {et~al.}(2013){Ribeiro}, {Lopes}, \& {Rembold}}]{ribeiro13}
{Ribeiro}, A.~L.~B., {Lopes}, P.~A.~A., \& {Rembold}, S.~B. 2013, \aap, 556,
  A74

\bibitem[{{Ribeiro} {et~al.}(2010){Ribeiro}, {Lopes}, \&
  {Trevisan}}]{ribeiro10}
{Ribeiro}, A.~L.~B., {Lopes}, P.~A.~A., \& {Trevisan}, M. 2010, \mnras, 409,
  L124

\bibitem[{{Rines} {et~al.}(2013){Rines}, {Geller}, {Diaferio}, \&
  {Kurtz}}]{rines13}
{Rines}, K., {Geller}, M.~J., {Diaferio}, A., \& {Kurtz}, M.~J. 2013, \apj,
  767, 15

\bibitem[{{Rines} {et~al.}(2003){Rines}, {Geller}, {Kurtz}, \&
  {Diaferio}}]{rines03}
{Rines}, K., {Geller}, M.~J., {Kurtz}, M.~J., \& {Diaferio}, A. 2003, \aj, 126,
  2152

\bibitem[{{Rosati} {et~al.}(2014){Rosati}, {Balestra}, {Grillo}, {Mercurio},
  {Nonino}, {Biviano}, {Girardi}, {Vanzella}, \& {Clash-VLT Team}}]{rosati14}
{Rosati}, P., {Balestra}, I., {Grillo}, C., {et~al.} 2014, The Messenger, 158,
  48

\bibitem[{{Rozo} {et~al.}(2010){Rozo}, {Wechsler}, {Rykoff}, {Annis}, {Becker},
  {Evrard}, {Frieman}, {Hansen}, {Hao}, {Johnston}, {Koester}, {McKay},
  {Sheldon}, \& {Weinberg}}]{rozo10}
{Rozo}, E., {Wechsler}, R.~H., {Rykoff}, E.~S., {et~al.} 2010, \apj, 708, 645

\bibitem[{{Rudnick} {et~al.}(2006){Rudnick}, {Labb{\'e}}, {F{\"o}rster
  Schreiber}, {Wuyts}, {Franx}, {Finlator}, {Kriek}, {Moorwood}, {Rix},
  {R{\"o}ttgering}, {Trujillo}, {van der Wel}, {van der Werf}, \& {van
  Dokkum}}]{rudnick06}
{Rudnick}, G., {Labb{\'e}}, I., {F{\"o}rster Schreiber}, N.~M., {et~al.} 2006,
  \apj, 650, 624

\bibitem[{{Rudnick} {et~al.}(2009){Rudnick}, {von der Linden}, {Pell{\'o}},
  {Arag{\'o}n-Salamanca}, {Marchesini}, {Clowe}, {De Lucia}, {Halliday},
  {Jablonka}, {Milvang-Jensen}, {Poggianti}, {Saglia}, {Simard}, {White}, \&
  {Zaritsky}}]{rudnick09}
{Rudnick}, G., {von der Linden}, A., {Pell{\'o}}, R., {et~al.} 2009, \apj, 700,
  1559

\bibitem[{{Ruel} {et~al.}(2014){Ruel}, {Bazin}, {Bayliss}, {Brodwin}, {Foley},
  {Stalder}, {Aird}, {Armstrong}, {Ashby}, {Bautz}, {Benson}, {Bleem},
  {Bocquet}, {Carlstrom}, {Chang}, {Chapman}, {Cho}, {Clocchiatti}, {Crawford},
  {Crites}, {de Haan}, {Desai}, {Dobbs}, {Dudley}, {Forman}, {George},
  {Gladders}, {Gonzalez}, {Halverson}, {Harrington}, {High}, {Holder},
  {Holzapfel}, {Hrubes}, {Jones}, {Joy}, {Keisler}, {Knox}, {Lee}, {Leitch},
  {Liu}, {Lueker}, {Luong-Van}, {Mantz}, {Marrone}, {McDonald}, {McMahon},
  {Mehl}, {Meyer}, {Mocanu}, {Mohr}, {Montroy}, {Murray}, {Natoli},
  {Nurgaliev}, {Padin}, {Plagge}, {Pryke}, {Reichardt}, {Rest}, {Ruhl},
  {Saliwanchik}, {Saro}, {Sayre}, {Schaffer}, {Shaw}, {Shirokoff}, {Song}, {{\v
  S}uhada}, {Spieler}, {Stanford}, {Staniszewski}, {Starsk}, {Story}, {Stubbs},
  {van Engelen}, {Vanderlinde}, {Vieira}, {Vikhlinin}, {Williamson}, {Zahn}, \&
  {Zenteno}}]{ruel14}
{Ruel}, J., {Bazin}, G., {Bayliss}, M., {et~al.} 2014, \apj, 792, 45

\bibitem[{{Sarazin}(1986)}]{sarazin86}
{Sarazin}, C.~L. 1986, Reviews of Modern Physics, 58, 1

\bibitem[{{Saro} {et~al.}(2013){Saro}, {Mohr}, {Bazin}, \& {Dolag}}]{saro13}
{Saro}, A., {Mohr}, J.~J., {Bazin}, G., \& {Dolag}, K. 2013, \apj, 772, 47

\bibitem[{{Sif{\'o}n} {et~al.}(2013){Sif{\'o}n}, {Menanteau}, {Hasselfield},
  {Marriage}, {Hughes}, {Barrientos}, {Gonz{\'a}lez}, {Infante}, {Addison},
  {Baker}, {Battaglia}, {Bond}, {Crichton}, {Das}, {Devlin}, {Dunkley},
  {D{\"u}nner}, {Gralla}, {Hajian}, {Hilton}, {Hincks}, {Kosowsky}, {Marsden},
  {Moodley}, {Niemack}, {Nolta}, {Page}, {Partridge}, {Reese}, {Sehgal},
  {Sievers}, {Spergel}, {Staggs}, {Thornton}, {Trac}, \& {Wollack}}]{sifon13}
{Sif{\'o}n}, C., {Menanteau}, F., {Hasselfield}, M., {et~al.} 2013, \apj, 772,
  25

\bibitem[{{Sif{\'o}n} {et~al.}(2016){Sif{\'o}n}, {Battaglia}, {Hasselfield},
  {Menanteau}, {Barrientos}, {Bond}, {Crichton}, {Devlin}, {D{\"u}nner},
  {Hilton}, {Hincks}, {Hlozek}, {Huffenberger}, {Hughes}, {Infante},
  {Kosowsky}, {Marsden}, {Marriage}, {Moodley}, {Niemack}, {Page}, {Spergel},
  {Staggs}, {Trac}, \& {Wollack}}]{sifon16}
{Sif{\'o}n}, C., {Battaglia}, N., {Hasselfield}, M., {et~al.} 2016, \mnras,
  461, 248

\bibitem[{{Sodre} {et~al.}(1989){Sodre}, {Capelato}, {Steiner}, \&
  {Mazure}}]{sodre89}
{Sodre}, Jr., L., {Capelato}, H.~V., {Steiner}, J.~E., \& {Mazure}, A. 1989,
  \aj, 97, 1279

\bibitem[{{Song} {et~al.}(2012){Song}, {Zenteno}, {Stalder}, {Desai}, {Bleem},
  {Aird}, {Armstrong}, {Ashby}, {Bayliss}, {Bazin}, {Benson}, {Bertin},
  {Brodwin}, {Carlstrom}, {Chang}, {Cho}, {Clocchiatti}, {Crawford}, {Crites},
  {de Haan}, {Dobbs}, {Dudley}, {Foley}, {George}, {Gettings}, {Gladders},
  {Gonzalez}, {Halverson}, {Harrington}, {High}, {Holder}, {Holzapfel},
  {Hoover}, {Hrubes}, {Joy}, {Keisler}, {Knox}, {Lee}, {Leitch}, {Liu},
  {Lueker}, {Luong-Van}, {Marrone}, {McDonald}, {McMahon}, {Mehl}, {Meyer},
  {Mocanu}, {Mohr}, {Montroy}, {Natoli}, {Nurgaliev}, {Padin}, {Plagge},
  {Pryke}, {Reichardt}, {Rest}, {Ruel}, {Ruhl}, {Saliwanchik}, {Saro}, {Sayre},
  {Schaffer}, {Shaw}, {Shirokoff}, {{\v S}uhada}, {Spieler}, {Stanford},
  {Staniszewski}, {Stark}, {Story}, {Stubbs}, {van Engelen}, {Vanderlinde},
  {Vieira}, {Williamson}, \& {Zahn}}]{song12}
{Song}, J., {Zenteno}, A., {Stalder}, B., {et~al.} 2012, \apj, 761, 22

\bibitem[{{Stalder} {et~al.}(2013){Stalder}, {Ruel}, {{\v S}uhada}, {Brodwin},
  {Aird}, {Andersson}, {Armstrong}, {Ashby}, {Bautz}, {Bayliss}, {Bazin},
  {Benson}, {Bleem}, {Carlstrom}, {Chang}, {Cho}, {Clocchiatti}, {Crawford},
  {Crites}, {de Haan}, {Desai}, {Dobbs}, {Dudley}, {Foley}, {Forman}, {George},
  {Gettings}, {Gladders}, {Gonzalez}, {Halverson}, {Harrington}, {High},
  {Holder}, {Holzapfel}, {Hoover}, {Hrubes}, {Jones}, {Joy}, {Keisler}, {Knox},
  {Lee}, {Leitch}, {Liu}, {Lueker}, {Luong-Van}, {Mantz}, {Marrone},
  {McDonald}, {McMahon}, {Mehl}, {Meyer}, {Mocanu}, {Mohr}, {Montroy},
  {Murray}, {Natoli}, {Nurgaliev}, {Padin}, {Plagge}, {Pryke}, {Reichardt},
  {Rest}, {Ruhl}, {Saliwanchik}, {Saro}, {Sayre}, {Schaffer}, {Shaw},
  {Shirokoff}, {Song}, {Spieler}, {Stanford}, {Staniszewski}, {Stark}, {Story},
  {Stubbs}, {van Engelen}, {Vanderlinde}, {Vieira}, {Vikhlinin}, {Williamson},
  {Zahn}, \& {Zenteno}}]{stalder13}
{Stalder}, B., {Ruel}, J., {{\v S}uhada}, R., {et~al.} 2013, \apj, 763, 93

\bibitem[{{Staniszewski} {et~al.}(2009){Staniszewski}, {Ade}, {Aird}, {Benson},
  {Bleem}, {Carlstrom}, {Chang}, {Cho}, {Crawford}, {Crites}, {de Haan},
  {Dobbs}, {Halverson}, {Holder}, {Holzapfel}, {Hrubes}, {Joy}, {Keisler},
  {Lanting}, {Lee}, {Leitch}, {Loehr}, {Lueker}, {McMahon}, {Mehl}, {Meyer},
  {Mohr}, {Montroy}, {Ngeow}, {Padin}, {Plagge}, {Pryke}, {Reichardt}, {Ruhl},
  {Schaffer}, {Shaw}, {Shirokoff}, {Spieler}, {Stalder}, {Stark},
  {Vanderlinde}, {Vieira}, {Zahn}, \& {Zenteno}}]{staniszewski09}
{Staniszewski}, Z., {Ade}, P.~A.~R., {Aird}, K.~A., {et~al.} 2009, \apj, 701,
  32

\bibitem[{Sunyaev \& Zel'dovich(1980)}]{sunyaev80}
Sunyaev, R., \& Zel'dovich, Y. 1980, ARAA, 18, 537

\bibitem[{{Sunyaev} \& {Zel'dovich}(1972)}]{sunyaev72}
{Sunyaev}, R.~A., \& {Zel'dovich}, Y.~B. 1972, Comments on Astrophysics and
  Space Physics, 4, 173

\bibitem[{{Tammann}(1972)}]{tammann72}
{Tammann}, G.~A. 1972, \aap, 21, 355

\bibitem[{{Vanderlinde} {et~al.}(2010){Vanderlinde}, {Crawford}, {de Haan},
  {Dudley}, {Shaw}, {Ade}, {Aird}, {Benson}, {Bleem}, {Brodwin}, {Carlstrom},
  {Chang}, {Crites}, {Desai}, {Dobbs}, {Foley}, {George}, {Gladders}, {Hall},
  {Halverson}, {High}, {Holder}, {Holzapfel}, {Hrubes}, {Joy}, {Keisler},
  {Knox}, {Lee}, {Leitch}, {Loehr}, {Lueker}, {Marrone}, {McMahon}, {Mehl},
  {Meyer}, {Mohr}, {Montroy}, {Ngeow}, {Padin}, {Plagge}, {Pryke}, {Reichardt},
  {Rest}, {Ruel}, {Ruhl}, {Schaffer}, {Shirokoff}, {Song}, {Spieler},
  {Stalder}, {Staniszewski}, {Stark}, {Stubbs}, {van Engelen}, {Vieira},
  {Williamson}, {Yang}, {Zahn}, \& {Zenteno}}]{vanderlinde10}
{Vanderlinde}, K., {Crawford}, T.~M., {de Haan}, T., {et~al.} 2010, \apj, 722,
  1180

\bibitem[{{von der Linden} {et~al.}(2014){von der Linden}, {Mantz}, {Allen},
  {Applegate}, {Kelly}, {Morris}, {Wright}, {Allen}, {Burchat}, {Burke},
  {Donovan}, \& {Ebeling}}]{vonderLinden14}
{von der Linden}, A., {Mantz}, A., {Allen}, S.~W., {et~al.} 2014, \mnras, 443,
  1973

\bibitem[{{White} {et~al.}(2010){White}, {Cohn}, \& {Smit}}]{white10}
{White}, M., {Cohn}, J.~D., \& {Smit}, R. 2010, \mnras, 408, 1818

\bibitem[{{Whitmore} {et~al.}(1993){Whitmore}, {Gilmore}, \&
  {Jones}}]{whitmore93}
{Whitmore}, B.~C., {Gilmore}, D.~M., \& {Jones}, C. 1993, \apj, 407, 489

\bibitem[{{Williamson} {et~al.}(2011){Williamson}, {Benson}, {High},
  {Vanderlinde}, {Ade}, {Aird}, {Andersson}, {Armstrong}, {Ashby}, {Bautz},
  {Bazin}, {Bertin}, {Bleem}, {Bonamente}, {Brodwin}, {Carlstrom}, {Chang},
  {Chapman}, {Clocchiatti}, {Crawford}, {Crites}, {de Haan}, {Desai}, {Dobbs},
  {Dudley}, {Fazio}, {Foley}, {Forman}, {Garmire}, {George}, {Gladders},
  {Gonzalez}, {Halverson}, {Holder}, {Holzapfel}, {Hoover}, {Hrubes}, {Jones},
  {Joy}, {Keisler}, {Knox}, {Lee}, {Leitch}, {Lueker}, {Luong-Van}, {Marrone},
  {McMahon}, {Mehl}, {Meyer}, {Mohr}, {Montroy}, {Murray}, {Padin}, {Plagge},
  {Pryke}, {Reichardt}, {Rest}, {Ruel}, {Ruhl}, {Saliwanchik}, {Saro},
  {Schaffer}, {Shaw}, {Shirokoff}, {Song}, {Spieler}, {Stalder}, {Stanford},
  {Staniszewski}, {Stark}, {Story}, {Stubbs}, {Vieira}, {Vikhlinin}, \&
  {Zenteno}}]{williamson11}
{Williamson}, R., {Benson}, B.~A., {High}, F.~W., {et~al.} 2011, \apj, 738, 139

\bibitem[{{Wu} {et~al.}(2013){Wu}, {Hahn}, {Evrard}, {Wechsler}, \&
  {Dolag}}]{wu13}
{Wu}, H.-Y., {Hahn}, O., {Evrard}, A.~E., {Wechsler}, R.~H., \& {Dolag}, K.
  2013, \mnras, 436, 460

\bibitem[{{York} {et~al.}(2000){York}, {Adelman}, {Anderson}, {Anderson},
  {Annis}, {Bahcall}, {Bakken}, {Barkhouser}, {Bastian}, {Berman}, {Boroski},
  {Bracker}, {Briegel}, {Briggs}, {Brinkmann}, {Brunner}, {Burles}, {Carey},
  {Carr}, {Castander}, {Chen}, {Colestock}, {Connolly}, {Crocker}, {Csabai},
  {Czarapata}, {Davis}, {Doi}, {Dombeck}, {Eisenstein}, {Ellman}, {Elms},
  {Evans}, {Fan}, {Federwitz}, {Fiscelli}, {Friedman}, {Frieman}, {Fukugita},
  {Gillespie}, {Gunn}, {Gurbani}, {de Haas}, {Haldeman}, {Harris}, {Hayes},
  {Heckman}, {Hennessy}, {Hindsley}, {Holm}, {Holmgren}, {Huang}, {Hull},
  {Husby}, {Ichikawa}, {Ichikawa}, {Ivezi{\'c}}, {Kent}, {Kim}, {Kinney},
  {Klaene}, {Kleinman}, {Kleinman}, {Knapp}, {Korienek}, {Kron}, {Kunszt},
  {Lamb}, {Lee}, {Leger}, {Limmongkol}, {Lindenmeyer}, {Long}, {Loomis},
  {Loveday}, {Lucinio}, {Lupton}, {MacKinnon}, {Mannery}, {Mantsch}, {Margon},
  {McGehee}, {McKay}, {Meiksin}, {Merelli}, {Monet}, {Munn}, {Narayanan},
  {Nash}, {Neilsen}, {Neswold}, {Newberg}, {Nichol}, {Nicinski}, {Nonino},
  {Okada}, {Okamura}, {Ostriker}, {Owen}, {Pauls}, {Peoples}, {Peterson},
  {Petravick}, {Pier}, {Pope}, {Pordes}, {Prosapio}, {Rechenmacher}, {Quinn},
  {Richards}, {Richmond}, {Rivetta}, {Rockosi}, {Ruthmansdorfer}, {Sandford},
  {Schlegel}, {Schneider}, {Sekiguchi}, {Sergey}, {Shimasaku}, {Siegmund},
  {Smee}, {Smith}, {Snedden}, {Stone}, {Stoughton}, {Strauss}, {Stubbs},
  {SubbaRao}, {Szalay}, {Szapudi}, {Szokoly}, {Thakar}, {Tremonti}, {Tucker},
  {Uomoto}, {Vanden Berk}, {Vogeley}, {Waddell}, {Wang}, {Watanabe},
  {Weinberg}, {Yanny}, {Yasuda}, \& {SDSS Collaboration}}]{york2000}
{York}, D.~G., {Adelman}, J., {Anderson}, Jr., J.~E., {et~al.} 2000, \aj, 120,
  1579

\bibitem[{{Zabludoff} \& {Franx}(1993)}]{zabludoff93}
{Zabludoff}, A.~I., \& {Franx}, M. 1993, \aj, 106, 1314

\bibitem[{{Zwicky}(1937)}]{zwicky37}
{Zwicky}, F. 1937, \apj, 86, 217

\end{thebibliography}

\end{document}